\documentclass[12pt]{article}
\usepackage{amssymb,amsmath,epsfig}

\renewcommand{\theequation}{\arabic{section}.\arabic{equation}}

\begin{document}

\title{\bf Hot Plasma Waves Surrounding the Schwarzschild Event Horizon in a Veselago Medium}
\author{M. Sharif \thanks{msharif@math.pu.edu.pk} and Noureen Mukhtar
\thanks{noureen.mukhtar@yahoo.com}\\
Department of Mathematics, University of the Punjab,\\
Quaid-e-Azam Campus, Lahore-54590, Pakistan.}
\date{}
\maketitle
\begin{abstract}
This paper investigates wave properties of hot plasma in a Veselago
medium. For the Schwarzschild black hole, the $3+1$ GRMHD equations
are re-formulated which are linearly perturbed and then Fourier
analyzed for rotating (non-magnetized and magnetized) plasmas. The
graphs of wave vector, refractive index and change in refractive are
used to discuss the wave properties. The results obtained confirm
the presence of Veselago medium for both rotating (non-magnetized
and magnetized) plasmas. This work generalized the isothermal plasma
waves in the Veselago medium to hot plasma case.
\end{abstract}
{\bf Keywords:} Veselago medium; $3+1$ formalism; GRMHD equations;
Isothermal plasma; Dispersion relations.\\
{\bf PACS:} 95.30.Sf; 95.30.Qd; 04.30.Nk

\section{Introduction}

Plasmas are found nearly everywhere in nature. These are
electrically conductive and give a strong respond to
electromagnetic fields. Plasma fills the interplanetary and
interstellar medium and are the components of the stars. To
explore the dynamics of magnetized plasma and characteristics of
black hole gravity when it acts in plasma's magnetic field, the
theory of general relativistic magnetohydrodynamics (GRMHD) is the
most accurate academic discipline. Schwarzschild black hole
has zero angular momentum so plasma present in the magnetosphere
moves only in the radial direction.

The gravity of black hole perturbs the magnetospheric plasma.
Relativists are always curious to study the effect of these
perturbations in the black hole regime. Regge and Wheeler \cite{1}
concluded that the Schwarzschild singularity remains stable when a
small non-spherical odd-parity perturbation is introduced. Zerilli
\cite{2} explored the same stability problem by considering an even
parity perturbation. The behavior of electric field generated by a
charged particle at rest near the Schwarzschild black hole was
discussed by Hanni and Ruffini \cite{3}. Sakai and Kawata \cite{4}
examined electron-positron plasma waves in the frame of two fluid
equations for the Schwarzschild black hole. Hirotoni and Tomimatsu
\cite{5} found that a small perturbation of polodial magnetic field
could highly disturb the plasma accretion. They assumed
non-stationary and axisymmetric perturbations of MHD accretion onto
the Schwarzschild magnetosphere. Zenginoglu et al. \cite{6} solved a
hyperboloidal initial value problem for the Bardeen-Press equation
to analyze the effect of gravitational perturbation on the
Schwarzschild spacetime.

The $3+1$ formalism (also called Arnowitt, Deser and Misner (ADM)
\cite{7}) is much helpful to study gravitational radiations from
black hole as well as analyzing the gravitational waves. To explore
the magnificent aspects of general relativity (GR), many authors
\cite{8}-\cite{10} adopted this technique. The electromagnetic
theory in the black hole regime was developed by Thorne and
Macdonald \cite{11,12}. Durrer and Straumann \cite{13} deduced some
basic results of GR by using this formalism. Holcomb and Tajima
\cite{14}, Holcomb \cite{15} and Dettmann et al. \cite{16} studied
the wave properties for the Friedmann universe. Buzzi et al.
\cite{17} investigated the plasma wave propagation close to the
Schwarzschild magnetosphere. Zhang \cite{18} composed the laws of
perfect GRMHD in $3+1$ formalism for a general spacetime. The same
author \cite{19} also described the role of cold plasma perturbation
in the vicinity of the Kerr black hole. Using this formulation,
Sharif and his collaborators \cite{20}-\cite{23} discussed plasma
(cold, isothermal and hot) wave properties with non-rotating as well
as rotating backgrounds.

Metamaterials are artificial materials that have unusual
electromagnetic properties and Veselago medium or Double negative
medium (DNG) is its most significant class. This medium has both
electric permittivity as well as magnetic permeability less than
zero. It is also known as negative refractive index medium (NIM) and
negative phase velocity medium (NPV). Many people
\cite{24}-\cite{28} considered this medium to explore the unusual
behavior of physical laws and their plausible applications.
Ziolkowski \cite{29} studied wave propagation in DNG medium both
analytically and numerically. Valanju et al. \cite{30} found
positive and very inhomogeneous wave refraction in this unusual
medium. Ramakrishna \cite{31} examined the problem of designing such
materials which have negative material parameters. He also discussed
the concept of perfect lens consisting of a slab of negative
refractive materials (NRM). Veselago \cite{32} verified the concepts
related to the energy, linear momentum and mass transferred by an
electromagnetic wave in a negative refraction medium. In a recent
paper \cite{33}, we have discussed the isothermal plasma wave
properties for the Schwarzschild magnetosphere in a Veselago medium.
The results verified the presence of this medium for only rotating
non-magnetized plasma.

In this paper, we investigate wave properties of hot plasma in the
vicinity of the Schwarzschild event horizon in a Veselago medium.
The format of the paper is as follows. In Section \textbf{2}, the
general line element in ADM $3+1$ formalism and its modification for
the Schwarzschild planar analogue is given. Linear perturbation and
Fourier analysis of the $3+1$ GRMHD equations for hot plasma is
provided in section \textbf{3}. Sections \textbf{4} and \textbf{5}
give the reduced form of the GRMHD equations for rotating
(non-magnetized and magnetized respectively) plasmas. In the last
section, summary of the results is given.

\section{3+1 Foliation and Planar Analogue of the Schwarzschild Spacetime}

The $3+1$ split of spacetime is an access to the field equations
in which four-dimensional spacetime is sliced into
three-dimensional spacelike hypersurfaces. In ADM $3+1$ formalism,
the general line element is \cite{19}
\begin{equation}\setcounter{equation}{1}\label{1}
ds^2=-\alpha^2dt^2+\gamma_{ij}(dx^i+\beta^idt)(dx^j+\beta^jdt),
\end{equation}
where the ratio of the fiducial proper time to universal time, i.e.,
$\frac{d\tau}{dt}$ is denoted by $\alpha$ (lapse function). When
FIDO (fiducial observer) changes his position from one hypersurface
to another, the shift vector $\beta^i$ calculates the change of
spatial coordinates. The components of three-dimensional
hypersurfaces are denoted by  $\gamma_{ij}~(i,j=1,2,3)$. A natural
observer associated with the above spacetime is known as FIDO. The
mathematical description of the Schwarzschild planar analogue is
given by
\begin{equation}\label{2}
ds^2=-\alpha^2(z)dt^2+dx^2+dy^2+dz^2,
\end{equation}
where the directions $z,~x$ and $y$ are analogous to the
Schwarzschild coordinates $r,~\phi$ and $\theta$ respectively. The
comparison of Eqs.(\ref{1}) and (\ref{2}) yields
\begin{equation}\label{3}
\alpha=\alpha(z),\quad\beta=0,\quad\gamma_{ij}=1~(i=j).
\end{equation}

\section{3+1 GRMHD Equations for the Schwarzschild Planar Analogue in a
Veselago Medium}

Appendix \textbf{A} contains the $3+1$ GRMHD equations in a
Veselagho medium for the plasma existing in the general line element
and the Schwarzschild planar analogue (Eqs.(\ref{1}) and (\ref{2})).
In the vicinity of the Schwarzschild magnetosphere, the specific
enthalpy for hot plasma is \cite{19}
\begin{eqnarray}\setcounter{equation}{1}\label{3}
\mu=\frac{\rho+p}{\rho_0},
\end{eqnarray}
where the rest mass density, moving mass density, pressure and
specific enthalpy are represented by $\rho_0,~\rho,~p$ and $\mu$
respectively. For cold plasma, we have $p=0$ while for isothermal
plasma, $p\neq0$ but specific enthalpy is constant. However,
specific enthalpy is variable for the hot plasma. This is the most
general plasma which reduces to cold and isothermal plasmas with
some restrictions. This equation shows the exchange of energy
between the plasma and fluid's magnetic field. It is obvious from
the above equation that for $\mu$ to be variable, $p$ must be
variable. In this paper, we have used the hot plasma along with the
Veselago medium in the vicinity of the Schwarzschild magnetosphere.
The $3+1$ GRMHD equations (Eqs.$(A10)$-$(A14)$) for hot plasma
surrounding the Schwarzschild event horizon become
\begin{eqnarray}
\label{4} &&\frac{\partial \textbf{B}}{\partial
t}=-\nabla\times(\alpha \textbf{V}\times \textbf{B}),\\
\label{5}&&\nabla.\textbf{B}=0,\\
\label{6} &&\frac{\partial (\rho+p) }{\partial t}+(\rho+p)\gamma^2
\textbf{V}. \frac{\partial \textbf{V}}{\partial t}+
(\rho+p)\gamma^2 V.(\alpha \textbf{V}.\nabla)
\textbf{V}\nonumber\\
&&+(\rho+p) \nabla.(\alpha\textbf{V})=0,
\end{eqnarray}
\begin{eqnarray} \label{7}
&&\left\{\left((\rho+p)\gamma^2+\frac{\textbf{B}^2}{4\pi}\right)\delta_{ij}
+(\rho+p)\gamma^4V_iV_j-\frac{1}{4\pi}B_iB_j\right\}
\left(\frac{1}{\alpha}\frac{\partial}{\partial
t}\right.\nonumber\\
&&\left.+\textbf{V}.\nabla\right)V^j+\gamma^2V_i(\textbf{V}.\nabla)(\rho+p)
-\left(\frac{\textbf{B}^2}{4\pi}\delta_{ij}-\frac{1}{4\pi}B_iB_j\right)V^j_{,k}V^k\nonumber\\
&&=-(\rho+p)\gamma^2a_i-p_{,i}+\frac{1}{4\pi}
(\textbf{V}\times\textbf{B})_i\nabla.(\textbf{V}\times\textbf{B})
-\frac{1}{8\pi\alpha^2}(\alpha\textbf{B})^2_{,i}\nonumber\\
&&+\frac{1}{4\pi\alpha}(\alpha B_i)_{,j}B^j-\frac{1}{4\pi\alpha}
[\textbf{B}\times\{\textbf{V}\times(\nabla\times(\alpha\textbf{V}\times\textbf{B}))\}]_i,\\
\label{8} &&(\frac{1}{\alpha}\frac{\partial}{\partial
t}+\textbf{V}.\nabla)(\rho+p)\gamma^2-\frac{1}{\alpha}\frac
{\partial p}{\partial t}+2(\rho+p)\gamma^2(\textbf{V}.\textbf{a})
+(\rho+p)\nonumber\\
&&\gamma^2(\nabla.\textbf{V})
-\frac{1}{4\pi\alpha}\left.(\textbf{V}\times\textbf{B}).(\textbf{V}\times\frac
{\partial \textbf{B}}{\partial t}\right.)
-\frac{1}{4\pi\alpha}\left.(\textbf{V}\times\textbf{B}).(\textbf{B}\times\frac{\partial
\textbf{B}}{\partial
t}\right.)\nonumber\\&&+\frac{1}{4\pi\alpha}\left(\textbf{V}\times\textbf{B}).
(\nabla\times\alpha\textbf{B}\right.)=0.
\end{eqnarray}
In rotating background, plasma is assumed to flow in two
dimensions, i.e., in $xz$-plane. Thus FIDO's measured magnetic
field $\textbf{B}$ and velocity $\textbf{V}$ become
\begin{eqnarray}\label{9}
\textbf{V}=V(z)\textbf{e}_x+u(z)\textbf{e}_z,\quad
\textbf{B}=B[\lambda(z)\textbf{e}_x+\textbf{e}_z],
\end{eqnarray}
where $B$ is an arbitrary constant. The quantities  $\lambda,~u$
and $V$ are related by \cite{20}
\begin{equation}\label{a}
V=\frac{V^F}{\alpha}+\lambda u,
\end{equation}
where $V^F$ is an integration constant. Thus the Lorentz factor
$\gamma=\frac{1}{\sqrt{1-\textbf{V}^2}}$ takes the form
\begin{equation}\label{b}
\gamma=\frac{1}{\sqrt{1-u^2-V^2}}.
\end{equation}

When the plasma flow is perturbed, the flow variables (mass
density $\rho$, pressure $p$, velocity $\textbf{V}$ and magnetic
field $\textbf{B}$) turn out to be
\begin{eqnarray}\label{10}
&&\rho=\rho^0+\delta\rho=\rho^0+\rho\widetilde{\rho},\quad
p=p^0+\delta p=p^0+p\widetilde{p},\nonumber\\
&&\textbf{V}=\textbf{V}^0+\delta\textbf{V}=\textbf{V}^0+\textbf{v},\quad
\textbf{B}=\textbf{B}^0+\delta\textbf{B}=\textbf{B}^0+B\textbf{b},
\end{eqnarray}
where unperturbed quantities are denoted by
$\rho^0,~p,~\textbf{V}^0,~\textbf{B}^0$ while $\delta\rho,~\delta
p,~\delta\textbf{V}$, $\delta\textbf{B}$ represent linearly
perturbed quantities. The following dimensionless quantities
$\widetilde{\rho},~\widetilde{p},~v_x,~v_z,~b_x$ and $b_z$ are
introduced for the perturbed quantities
\begin{eqnarray}\label{11}
&&\tilde{\rho}=\tilde{\rho}(t,z),\quad
\tilde{p}=\tilde{p}(t,z),\quad\textbf{v}=\delta\textbf{V}=v_x(t,z)\textbf{e}_x
+v_z(t,z)\textbf{e}_z,\nonumber\\
&&\textbf{b}=\frac{\delta\textbf{B}}{B}=b_x(t,z)\textbf{e}_x
+b_z(t,z)\textbf{e}_z.
\end{eqnarray}
When we insert these linear perturbations in the perfect GRMHD
equations (Eqs.(\ref{4})-(\ref{8})), it follows that
\begin{eqnarray}\label{12}
&&\frac{\partial(\delta\textbf{B})}{\partial
t}=-\nabla\times(\alpha\textbf{v}\times\textbf{B})
-\nabla\times(\alpha\textbf{V}\times\delta\textbf{B}),\\\label{13}
&&\nabla.(\delta\textbf{B})=0,\\\label{14}
&&\frac{1}{\alpha}\frac{\partial(\delta\rho+\delta
p)}{\partial
t}+(\rho+p)\gamma^2\textbf{V}.(\frac{1}{\alpha}\frac{\partial}{\partial
t}+\textbf{V}.\nabla)\textbf{v}\nonumber+(\rho+p)(\nabla.\textbf{v})\nonumber\\
&&=-2(\rho+p)\gamma^2(\textbf{V}.\textbf{v})(\textbf{V}.\nabla)\ln\gamma
-(\rho+p)\gamma^2(\textbf{V}.\nabla\textbf{V}).\textbf{v}\nonumber\\
&&+(\rho+p)(\textbf{v}.\nabla\ln u),\\\label{15}
&&\left\{\left((\rho+p)\gamma^2
+\frac{\textbf{B}^2}{4\pi}\right)\delta_{ij}+(\rho+p)\gamma^4V_iV_j
-\frac{1}{4\pi}B_iB_j\right\}\frac{1}{\alpha}\frac{\partial
v^j}{\partial t}\nonumber\\
&&+\frac{1}{4\pi}[\textbf{B}\times\{\textbf{V}
\times\frac{1}{\alpha}\frac{\partial(\delta\textbf{B})}{\partial
t}\}]_i+(\rho+p)\gamma^2v_{i,j}V^j+(\rho+p)\nonumber\\
&&\times\gamma^4V_iv_{j,k}V^jV^k
+\gamma^2V_i(\textbf{V}.\nabla)(\delta \rho+\delta
p)+\gamma^2V_i(\textbf{v}.\nabla)(\rho+p)\nonumber\\
&&+\gamma^2v_i(\textbf{V}.\nabla)(\rho+p)
+\gamma^4(2\textbf{V}.\textbf{v})V_i(\textbf{V}.\nabla)(\rho+p)
-\frac{1}{4\pi\alpha}\{(\alpha\delta B_i)_{,j}\nonumber\\
&&-(\alpha\delta B_j)_{,i}\}B^j=-(\delta
p)_i-\gamma^2\{(\delta\rho+\delta
p)+2(\rho+p)\gamma^2(\textbf{V}.\textbf{v})\}a_i\nonumber\\
&&+\frac{1}{4\pi\alpha}\{(\alpha B_i)_{,j}-(\alpha B_
j)_{,i}\}\delta
B^j-(\rho+p)\gamma^4(v_iV^j+v^jV_i)V_{k,j}V^k\nonumber\\
&&-\gamma^2\{(\delta\rho+\delta
p)V^j+2(\rho+p)\gamma^2(\textbf{V}.\textbf{v})V^j+(\rho+p)v^j\}V_{i,j}\nonumber\\
&&-\gamma^4V_i\{(\delta\rho+\delta
p)V^j+4(\rho+p)\gamma^2(\textbf{V}.\textbf{v})V^j+(\rho+p)v^j\}V_{j,k}V^k,\\\label{16}
&&\gamma^2\frac{1}{\alpha}\frac{\partial(\delta\rho+\delta
p)}{\partial
t}+\textbf{v}.\nabla(\rho+p)\gamma^2-\frac{1}{\alpha}\frac{\partial(\delta
p)}{\partial t}+(\textbf{V}.\nabla)(\delta\rho+\delta
p)\gamma^2\nonumber\\
&&+2(\rho+p)\gamma^4(\textbf{V}.\nabla)(\textbf{V}.\textbf{v})
+2(\rho+p)\gamma^2(\textbf{v}.\textbf{a})
+4(\rho+p)\gamma^4(\textbf{V}.\textbf{v})\nonumber\\
&&(\textbf{V}.\textbf{a})+2(\delta\rho+\delta
p)\gamma^2(\textbf{V}.\textbf{a})+(\rho+p)\gamma^2(\nabla.\textbf{v})+2(\rho+p)\gamma^4\nonumber\\
&&(\textbf{V}.\textbf{v})(\nabla.\textbf{V})+(\delta\rho+\delta
p)\gamma^2(\nabla.\textbf{V})=\frac{1}{4\pi\alpha}
[\textbf{v}.(\textbf{B}.\frac{\partial\textbf{B}}{\partial
t})\textbf{V}+\textbf{V}.(\textbf{B}.\frac{\partial\textbf{B}}{\partial
t})\textbf{v}\nonumber
\end{eqnarray}
\begin{eqnarray}
&&+\textbf{V}.(\textbf{B}.\delta\textbf{B})\textbf{V}
+\textbf{V}.(\delta\textbf{B}\frac{\partial\textbf{B}}{\partial
t})\textbf{V}-\textbf{v}.(\textbf{B}.\textbf{V})\frac{\partial\textbf{B}}{\partial
t}-\textbf{V}.(\textbf{B}.\textbf{V})\frac{\partial\delta\textbf{B}}
{\partial t}\nonumber\\
&&-\textbf{V}.(\textbf{B}.\textbf{v})\frac{\partial\delta\textbf{B}}{\partial
t}-\textbf{V}.(\delta\textbf{B}.\textbf{V})\frac{\partial\textbf{B}}{\partial
t}]-\frac{1}{4\pi\alpha}[\textbf{V}.(\textbf{B}.\textbf{B})\frac{\partial\textbf{v}}{\partial
t}-\textbf{V}.(\textbf{B}.\frac{\partial\delta\textbf{v}}{\partial
t})\textbf{B}]\nonumber\\
&&+\frac{1}{4\pi}[(\textbf{v}\times\textbf{B}+\textbf{V}\times\delta\textbf{B})
.(\nabla\times\textbf{B})+(\textbf{V}\times\textbf{B}).(\nabla\times\delta\textbf{B})]
\end{eqnarray}

The component form of these equations, using Eq.(\ref{11}), become
\begin{eqnarray}
\label{17}&&\frac{1}{\alpha}\frac{\partial b_x}{\partial
t}-ub_{x,z}=(ub_x-Vb_z-v_x+\lambda v_z)\nabla
\ln\alpha\nonumber\\
&&-(v_{x,z}-\lambda
v_{z,z}-\lambda'v_z+V'b_z+Vb_{z,z}-u'b_x),\\\label{18}
&&\frac{1}{\alpha}\frac{\partial b_z}{\partial t}=0,\\\label{19}
&&b_{z,z}=0,\\ \label{20}
&&\rho\frac{1}{\alpha}\frac{\partial\tilde{\rho}}{\partial t}
+p\frac{1}{\alpha}\frac{\partial\tilde{p}}{\partial
t}+(\rho+p)\gamma^2V(\frac{1}{\alpha}\frac{\partial{v_x}}{\partial
t}+uv_{x,z})+(\rho+p)\gamma^2u\nonumber\\
&&\times\frac{1}{\alpha}\frac{\partial{v_z}}{\partial
t}+(\rho+p)(1+\gamma^2u^2)v_{z,z}=-\gamma^2u(\rho+p)[(1+2\gamma^2V^2)V'\nonumber\\
&&+2\gamma^2uVu']v_x+(\rho+p)[(1-2\gamma^2u^2)(1+\gamma^2u^2)\frac{u'}{u}\nonumber\\
&&-2\gamma^4u^2VV']v_z,\\
\label{21}&&\left\{(\rho+p)\gamma^2(1+\gamma^2V^2)
+\frac{B^2}{4\pi}\right\}\frac{1}{\alpha}\frac{\partial
v_x}{\partial t}+\left\{(\rho+p)\gamma^4uV-\frac{\lambda B
^2}{4\pi}\right\}\nonumber\\
&&\times\frac{1}{\alpha}\frac{\partial v_z}{\partial
t}+\left\{(\rho+p)\gamma^2(1+\gamma^2V^2)
+\frac{B^2}{4\pi}\right\}uv_{x,z}+\left\{(\rho+p)\gamma^4uV\right.\nonumber\\
&&\left.-\frac{\lambda B^2}{4\pi}\right\}uv_{z,z}
-\frac{B^2}{4\pi}(1+u^2)b_{x,z}-\frac{B^2}{4\pi\alpha}\left\{\alpha'(1+u^2)+\alpha
uu'\right\}b_x\nonumber\\
&&+\gamma^2u(\rho\tilde{\rho}+p\tilde{p})\left\{(1+\gamma^2V^2)V'+\gamma^2uVu'\right\}
+\gamma^2uV(\rho'\tilde{\rho}+\rho\tilde{\rho}'\nonumber\\
&&+p'\tilde{p}+p\tilde{p}')+[(\rho+p)\gamma^4u
\left\{(1+4\gamma^2V^2)uu'+4VV'(1+\gamma^2V^2)\right\}\nonumber\\
&&+\frac{B^2u\alpha'}{4\pi\alpha}
+\gamma^2u(1+2\gamma^2V^2)(\rho'+p')]v_x
+[(\rho+p)\gamma^2\left\{(1+2\gamma^2u^2)\right.\nonumber\\
&&\left.(1+2\gamma^2V^2)V'-\gamma^2V^2V'
+2\gamma^2(1+2\gamma^2u^2)uVu'\right\}-\frac{B^2u}
{4\pi\alpha}(\lambda\alpha)'\nonumber\\
&&+\gamma^2V(1+2\gamma^2u^2)(\rho'+p')]v_z=0, \label{22}
\end{eqnarray}
\begin{eqnarray}
&&\left\{(\rho+p)\gamma^2(1+\gamma^2u^2)
+\frac{\lambda^2B^2}{4\pi}\right\}\frac{1}{\alpha}\frac{\partial
v_z}{\partial t}+\left\{(\rho+p)\gamma^4uV -\frac{\lambda B
^2}{4\pi}\right\}\nonumber\\
&&\times\frac{1}{\alpha}\frac{\partial v_x}{\partial t}
+\left\{(\rho+p)\gamma^2(1+\gamma^2u^2)+\frac{\lambda^2B^2}{4\pi}\right\}
uv_{z,z}+\left\{(\rho+p)\gamma^4uV \right.\nonumber\\
&&\left.-\frac{\lambda B^2}{4\pi}\right\}uv_{x,z}+\frac{\lambda
B^2}{4\pi}(1+u^2)b_{x,z}+\frac{B^2}{4\pi\alpha}\left\{(\alpha\lambda)'
-\alpha'\lambda+u\lambda(u\alpha'\right.\nonumber\\
&&\left.+u'\alpha)\right\}b_x+(\rho\tilde{\rho}+p\tilde{p})\gamma^2\left\{a_z
+uu'(1+\gamma^2u^2)+\gamma^2u^2VV'\right\}\nonumber\\
&&+(1+\gamma^2u^2)(p'\tilde{p}+p\tilde{p}')
+\gamma^2u^2(\rho'\tilde{\rho}+\rho\tilde{\rho}')+[(\rho+p)\gamma^4\nonumber\\
&&\times\{u^2V'(1+4\gamma^2V^2)+2V(a_z+uu'(1+2\gamma^2u^2))\}-\frac{\lambda
B^2u\alpha'}{4\pi\alpha}\nonumber\\
&&+2\gamma^4u^2V(\rho'+p')]v_x+[(\rho+p)\gamma^2
\left\{u'(1+\gamma^2u^2)(1+4\gamma^2u^2)\right.\nonumber\\
&&\left.+2u\gamma^2(a_z+(1+2\gamma^2u^2)VV')\right\}
+\frac{\lambda B^2u}{4\pi\alpha}(\alpha\lambda)'
+2\gamma^2u(1\nonumber\\
&&+\gamma^2u^2)(\rho'+p')]v_z=0,\\
\label{23} &&\frac{1}{\alpha}\gamma^2\rho\frac{\partial
\tilde{\rho}}{\partial t}+\frac{1}{\alpha}\gamma^2p\frac{\partial
\tilde{p}}{\partial
t}+\gamma^2(\rho'+p')v_z+u\gamma^2(\rho\tilde{\rho}_{,z}+p\tilde{p}_{,z}+\rho'\tilde{\rho}\nonumber\\
&&+p'\tilde{p})-\frac{1}{\alpha}p\frac{\partial
\tilde{p}}{\partial
t}+2\gamma^2u(\rho\tilde{\rho}+p\tilde{p})a_z+\gamma^2u'(\rho\tilde{\rho}+p\tilde{p})+2(\rho\nonumber\\
&&+p)\gamma^4(uV'+2uVa_z+u'V)v_x+2(\rho+p)\gamma^2(2\gamma^2uu'+a_z\gamma^4\nonumber\\
&&+2\gamma^2u^2a_z)v_z+2(\rho+p)\gamma^4uVv_{x,z}+(\rho+p)\gamma^2(1+2\gamma^2u^2)\nonumber\\
&&\times v_{z,z}-\frac{B^2}{4\pi\alpha}[(V^2+u^2)\lambda
b_x+(V^2+u^2)b_z-\lambda V(\lambda V\nonumber\\
&&+u)\frac{\partial b_x}{\partial t}-u(\lambda V+u)\frac{\partial
b_z}{\partial t}]-\frac{B^2}{4\pi\alpha}[(V-\lambda
u)v_{x,t}+\lambda(u\lambda\nonumber\\
&&-V)v_{z,t}]+\frac{B^2}
{4\pi}(\lambda\lambda'v_z-\lambda'v_x-\lambda'Vb_z\nonumber\\
&&+\lambda'ub_x-V b_{x,z}+u\lambda b_{x,z})=0.
\end{eqnarray}
For the purpose of Fourier analysis, the following harmonic spacetime dependence
of perturbation is assumed
\begin{eqnarray}\label{24}
\widetilde{\rho}(t,z)=c_1e^{-\iota(\omega t-kz)},&\quad&
\widetilde{p}(t,z)=c_2e^{-\iota(\omega t-kz)},\nonumber\\
v_z(t,z)=c_3e^{-\iota(\omega t-kz)},&\quad&
v_x(t,z)=c_4e^{-\iota(\omega t-kz)},\nonumber\\
b_z(t,z)=c_5e^{-\iota(\omega t-kz)},&\quad&
b_x(t,z)=c_6e^{-\iota(\omega t-kz)}.
\end{eqnarray}
Here $k$ and $\omega$ are the $z$-component of the wave vector
$(0,0,k)$ and angular frequency respectively. Plasma wave properties
near the event horizon can be explored by the wave vector which is
also used to obtain refractive index. We define wave vector and
refractive index as follows:
\begin{itemize}
\item\textbf{Wave Vector}: The direction in which a plane wave
propagates is represented by a wave vector. Its magnitude gives the
wave number.
\item\textbf{Refractive Index}: When light travels from one
medium to another (usually from vacuum) then its ratio between the
two mediums is given by the refractive index. The change in the
refractive index with respect to angular frequency decides whether
the dispersion will be normal or anomalous.
\end{itemize}
Using Eq.(\ref{24}) in Eqs.(\ref{17})-(\ref{23}), we get their
Fourier analyzed form
\begin{eqnarray}\label{25}
&&c_{4}(\alpha'+\iota k\alpha)-c_3\
\left\{(\alpha\lambda)'+\iota k\alpha\lambda\ \right\}-c_5(\alpha
V)'-c_6\{(\alpha
u)'+\iota\omega\nonumber\\
&&+\iota ku\alpha\}=0,\\\label{26}
&&c_5(\frac{-\iota\omega}{\alpha})=0,\\\label{27} &&c_5\iota
k=0,\\\label{28}&&c_1(\frac{-\iota\omega}{\alpha}\rho)+c_2(\frac{-\iota\omega}{\alpha}p)
+c_3(\rho+p)[\frac{-\iota\omega}{\alpha}\gamma^2u+(1+\gamma^2u^2)\iota k\nonumber\\
&&-(1-2\gamma^2u^2)(1+\gamma^2u^2)\frac{u'}{u}+2\gamma^4u^2VV']
+c_4(\rho+p)\gamma^2[(\frac{-\iota\omega}{\alpha}\nonumber\\
&&+\iota ku)V+u(1+2\gamma^2V^2)V'+2\gamma^2u^2Vu']=0,\\
\label{29}&&c_1[\rho\gamma^2u\{(1+\gamma^2V^2)V'+\gamma^2Vuu'\}+\gamma^2Vu(\rho'+\iota
k\rho)]\nonumber\\
&&+c_2[p\gamma^2u\{(1+\gamma^2V^2)V'+\gamma^2Vuu'\}+\gamma^2Vu(p'+\iota kp)]\nonumber\\
&&+c_3[(\rho+p)\gamma^2
\{(1+2\gamma^2u^2)(1+2\gamma^2V^2)V'+(\frac{-\iota\omega}{\alpha}+\iota
ku)\gamma^2Vu\nonumber\\
&&-\gamma^2V^2V'+2\gamma^2(1+2\gamma^2u^2)uVu'\}+\gamma^2V(1+2\gamma^2u^2)(\rho'+p')\nonumber\\
&&-\frac{B^2u}{4\pi\alpha}(\lambda\alpha)'+\frac{\lambda
B^2}{4\pi}(\frac{\iota\omega}{\alpha}-\iota
ku)]+c_4[(\rho+p)\gamma^4u\{(1+4\gamma^2V^2)\nonumber\\ &&\times
uu'+4VV'(1+\gamma^2V^2)\}
+(\rho+p)\gamma^2(1+\gamma^2V^2)(\frac{-\iota\omega}{\alpha}+\iota
ku)\nonumber\\
&&+\gamma^2u(1+2\gamma^2V^2)(\rho'+p')+\frac{B^2u\alpha'}{4\pi\alpha}
-\frac{B^2}{4\pi}(\frac{\iota\omega}{\alpha}-\iota
ku)]\nonumber\\
&&-c_6\frac{B^2}{4\pi\alpha}[\alpha uu'+\alpha'(1+u^2)+(1+u^2)\iota
k\alpha]=0,\label{30}
\end{eqnarray}
\begin{eqnarray}
&&c_1[\rho\gamma^2\{a_z+(1+\gamma^2u^2)uu'+\gamma^2u^2VV'\}+\gamma^2u^2(\rho'+\iota
k\rho)]\nonumber\\
&&+c_2[p\gamma^2\{a_z+(1+\gamma^2u^2)uu'+\gamma^2u^2VV'\}+(1+\gamma^2u^2)\nonumber\\
&&\times(p'+\iota kp)]+c_3[(\rho+p)\gamma^2
\{(1+\gamma^2u^2)(\frac{-\iota\omega}{\alpha}+\iota ku)\nonumber\\
&&+u'(1+\gamma^2u^2)(1+4\gamma^2u^2)+2u\gamma^2(a_z
+(1+2\gamma^2u^2)VV')\}\nonumber\\
&&+2\gamma^2u(1+\gamma^2u^2)(\rho'+p')+\frac{\lambda
B^2u}{4\pi\alpha}(\lambda\alpha)'-\frac{\lambda^2
B^2}{4\pi}(\frac{\iota\omega}{\alpha}-\iota ku)]\nonumber\\
&&+c_4[(\rho+p)\gamma^4\{(\frac{-\iota\omega}{\alpha}+\iota
ku)uV+u^2V'(1+4\gamma^2V^2)+2V(a_z\nonumber\\
&&+(1+2\gamma^2u^2)uu')\}+2\gamma^4u^2V(\rho'+p')+\frac{\lambda
B^2}{4\pi}(\frac{\iota\omega}{\alpha}-\iota ku)\nonumber\\
&&-\frac{\lambda B^2u\alpha'}{4\pi\alpha}]
+c_6[\frac{B^2}{4\pi\alpha}\{-(\lambda\alpha)'
+\alpha'\lambda-u\lambda(u\alpha'+u'\alpha)\}\nonumber\\
&&+\frac{\lambda B^2}{4\pi}(1+u^2)\iota k]=0,\\
\label{31}&&c_1\{(\frac{-\iota\omega}{\alpha}\gamma^2+\iota ku
\gamma^2+2u\gamma^2a_z+\gamma^2u')\rho+u\rho'\gamma^2\}
+c_2\{(\frac{\iota\omega}{\alpha}(1-\gamma^2)\nonumber\\
&&+\iota ku\gamma^2+2\gamma^2ua_z+\gamma^2u')p+u\gamma^2p'\}+c_3\gamma^2\{(\rho'+p')+2\nonumber\\
&&\times(2\gamma^4uu'+a_z+2\gamma^2u^2a_z)(\rho+p)+(1+2\gamma^2u^2)(\rho+p)\iota
k+\frac{\lambda B^2}{4\pi\alpha}\nonumber\\
&&\times(\lambda u-V)\iota\omega+\alpha\lambda'\}+c_4[2(\rho+p)\gamma^2\{(u
V'+2uVa_z+u'V)+uV\iota
k\}\nonumber\\
&&+\frac{B^2}{4\pi\alpha}(V-u\lambda)\iota\omega-\alpha\lambda']
+c_6[\frac{-B^2}{4\pi\alpha}\{(V^2+u^2)\lambda+\lambda
V(\lambda V+u)\iota\omega\}\nonumber\\
&&-\alpha\lambda'u+\iota k\alpha(V-u\lambda)]=0.
\end{eqnarray}
Dispersion relations will be obtained by using these equations.

\section{Rotating Non-Magnetized Flow with Hot Plasma}

In this section, rotating non-magnetized background of plasma flow
is assumed, i.e., $\textbf{B}=0$. Thus, the evolution equations
(\ref{4}) and (\ref{5}) of magnetic field are satisfied.
Substituting $B=0=\lambda$ and $c_5=0=c_6$ in the Fourier analyzed
perturbed GRMHD equations (Eqs.(\ref{28})-(\ref{31})), we have
\begin{eqnarray}{\setcounter{equation}{1}}
\label{32}&&c_1(\frac{-\iota\omega}{\alpha}\rho)+c_2(\frac{-\iota\omega}{\alpha}p)
+c_3(\rho+p)[\frac{-\iota\omega}{\alpha}\gamma^2u+(1+\gamma^2u^2)\iota k\nonumber\\
&&-(1-2\gamma^2u^2)(1+\gamma^2u^2)\frac{u'}{u}+2\gamma^4u^2VV']
+c_4(\rho+p)\gamma^2[(\frac{-\iota\omega}{\alpha}\nonumber\\
&&+\iota ku)V +u(1+2\gamma^2V^2)V'+2\gamma^2u^2Vu']=0,\\
\label{33}&&c_1[\rho\gamma^2
u\{(1+\gamma^2V^2)V'+\gamma^2Vuu'\}+\gamma^2Vu(\rho'+\iota
k\rho)]+c_2[p\gamma^2u\nonumber\\
&&\times\{(1+\gamma^2V^2)V'+\gamma^2Vuu'\}+\gamma^2Vu(p'+\iota
kp)]+c_3[(\rho+p)\gamma^2\nonumber\\
&&\times\{(\frac{-\iota\omega}{\alpha}+\iota
ku)\gamma^2Vu+(1+2\gamma^2u^2)(1+2\gamma^2V^2)V'-\gamma^2V^2V'\nonumber\\
&&+2\gamma^2(1+2\gamma^2u^2)uVu'\}
+\gamma^2V(1+2\gamma^2u^2)(\rho'+p')]+c_4[(\rho+p)\nonumber\\
&&\{\gamma^2(1+\gamma^2V^2)(\frac{-\iota\omega}{\alpha}+\iota
ku)+\gamma^4u((1+4\gamma^2V^2)uu'+4VV'(1+\nonumber\\
&&\gamma^2V^2))\} +\gamma^2u(1+2\gamma^2V^2)(\rho'+p')]=0,\\
\label{34}&&c_1[\rho\gamma^2\{a_z+(1+\gamma^2u^2)uu'
+\gamma^2u^2VV'\}+\gamma^2u^2(\rho'+\iota
k\rho)]\nonumber\\
&&+c_2[p\gamma^2\{a_z
+(1+\gamma^2u^2)uu'+\gamma^2u^2VV'\}+(p'+\iota
kp)\times\nonumber\\
&&(1+\gamma^2u^2)]+c_3[(\rho+p)\gamma^2\{(1+\gamma^2u^2)(\frac{-\iota\omega}{\alpha}+\iota
ku)+u'(1+\gamma^2u^2)\nonumber\\
&&\times(1+4\gamma^2u^2)+2u\gamma^2(a_z+(1+2\gamma^2u^2)VV')\}
+2\gamma^2u(1+\gamma^2u^2)\nonumber\\
&&\times(\rho'+p')]+c_4[(\rho+p)\gamma^4\{(\frac{-\iota\omega}{\alpha}
+\iota ku)uV+u^2V'(1+4\gamma^2V^2)\nonumber\\
&&+2V(a_z+(1+2\gamma^2u^2)uu')\}+2\gamma^4u^2V(\rho'+p')]=0,\\\label{35}
&&c_1\{(\frac{-\iota\omega}{\alpha}\gamma^2+\iota ku
\gamma^2+2u\gamma^2a_z+\gamma^2u')\rho+u\rho'\gamma^2\}
+c_2\{(\frac{\iota\omega}{\alpha}(1-\gamma^2)\nonumber\\
&&+\iota
ku\gamma^2+2\gamma^2ua_z+\gamma^2u')p+u\gamma^2p'\}+c_3\gamma^2\{(\rho'+p')+2
(2\gamma^2uu'\nonumber\\
&&+a_z+2\gamma^2u^2a_z)(\rho+p)+(1+2\gamma^2u^2)(\rho+p)\iota
k\}+c_4(\rho+p)2\gamma^4\nonumber\\
&&\{(u V'+2uVa_z+u'V)+uV\iota k\}=0.
\end{eqnarray}

\subsection{Numerical Solutions}

For the rotating non-magnetized plasma, we use the following assumptions
to find out the numerical solutions
\begin{itemize}
\item Specific enthalpy: $\mu=\sqrt{\frac{1-(\tanh(10z)/10)^2}{2}},$
\item Time lapse: $\alpha=\tanh(10z)/10,$
\item Stationary fluid: $\alpha\gamma=1$ with velocity components $V=u$
gives the following relation:
$\alpha\gamma=1~\Rightarrow\gamma=1/\sqrt{1-u^2-V^2}=1/\alpha$.
\item  Velocity components: $u=V,~x$ and $z$-components of velocity
yield $u=V=-\sqrt\frac{1-\alpha^2}{2}$.
\item Stiff fluid: $\rho=p=\mu/2$.
\end{itemize}
These assumptions satisfy the GRMHD equations
(Eqs.(\ref{4})-(\ref{8})) for the region $1.4\leq z\leq10,
0\leq\omega\leq 10$. A complex dispersion relation \cite{34} is
obtained by solving the determinant of the coefficients of
constants of Eqs.(\ref{32})-(\ref{35}). We obtain a quartic equation in $k$
from the real part of the determinant
\begin{equation}\label{36}
A_1(z)k^4+A_2(z,\omega)k^3+A_3(z,\omega)k^2+A_4(z,\omega)k+A_5(z,\omega)=0
\end{equation}
which yields four values of $k$ out of which two are real and two
complex conjugate. The imaginary part gives a cubic
equation in $k$
\begin{equation}\label{37}
B_1(z)k^3+B_2(z,\omega)k^2+B_3(z,\omega)k+B_4(z,\omega)=0
\end{equation}
from which one real value of $k$ is obtained and the remaining two
are complex conjugate of each other. From the real values of $k$
((\ref{36}) and (\ref{37})), wave vector, refractive index and its
change with respect to angular frequency are shown in Figures
\textbf{1-3}.
\begin{figure}
\center \epsfig{file=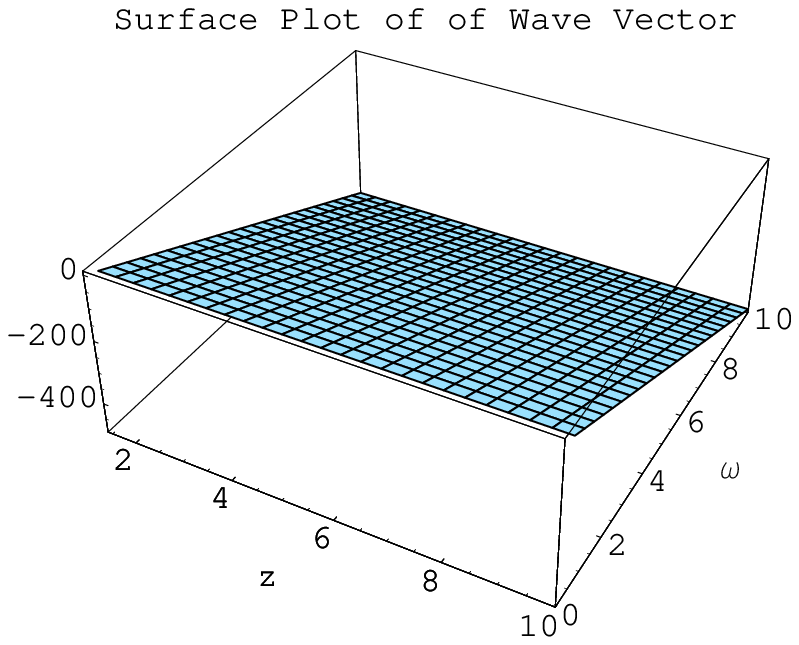,width=0.40\linewidth} \center
\begin{tabular}{cc}
\epsfig{file=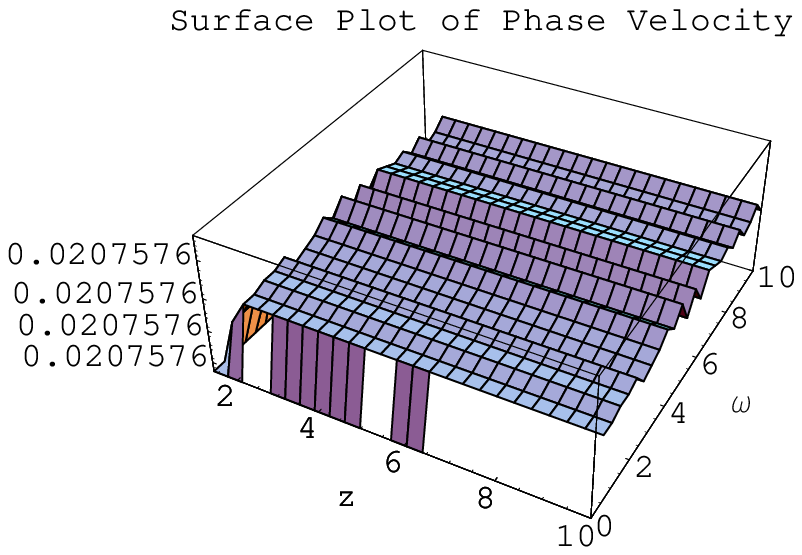,width=0.45\linewidth}
\epsfig{file=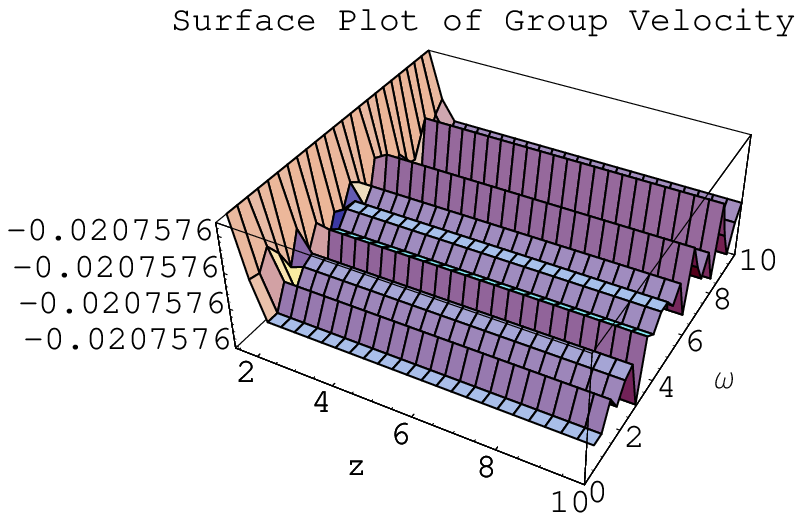,width=0.45\linewidth}\\
\epsfig{file=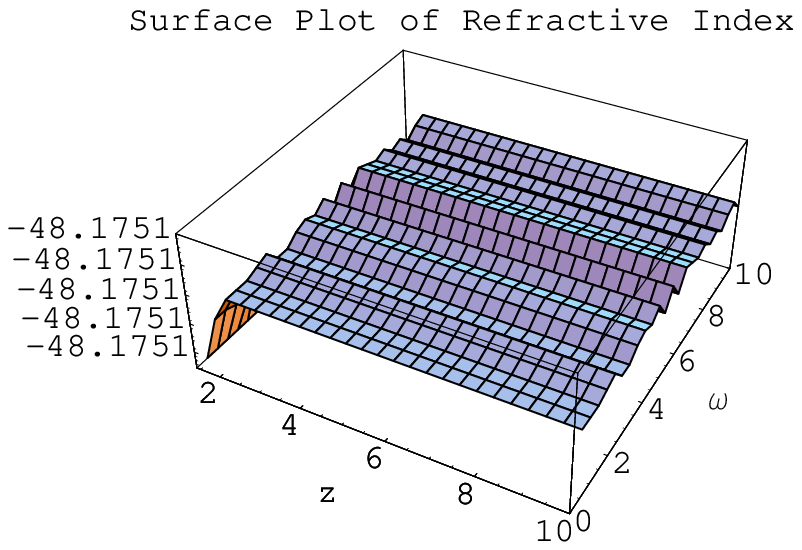,width=0.40\linewidth}
\epsfig{file=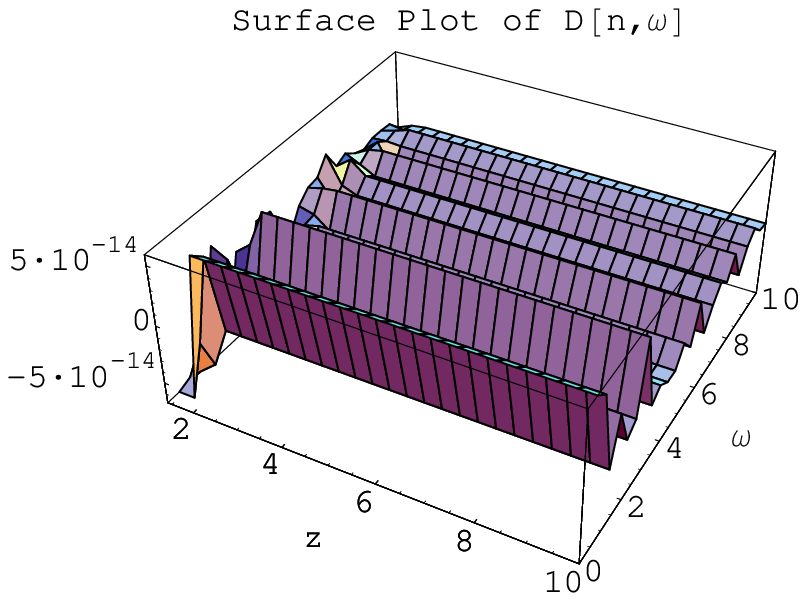,width=0.35\linewidth}
\end{tabular}
\caption{Waves are directed towards the event horizon. The
dispersion is found to be normal as well as anomalous randomly.}
\end{figure}
\begin{figure} \center
\epsfig{file=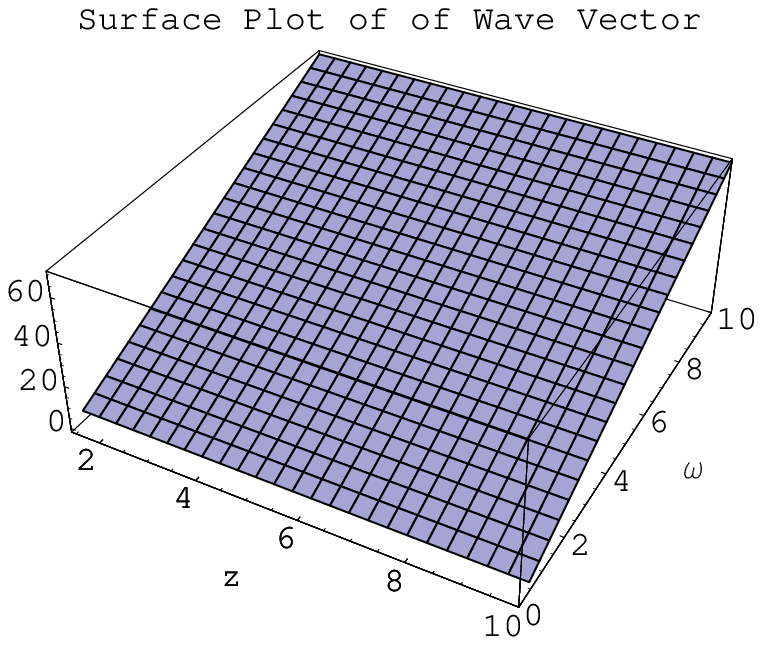,width=0.40\linewidth} \center
\begin{tabular}{cc}
\epsfig{file=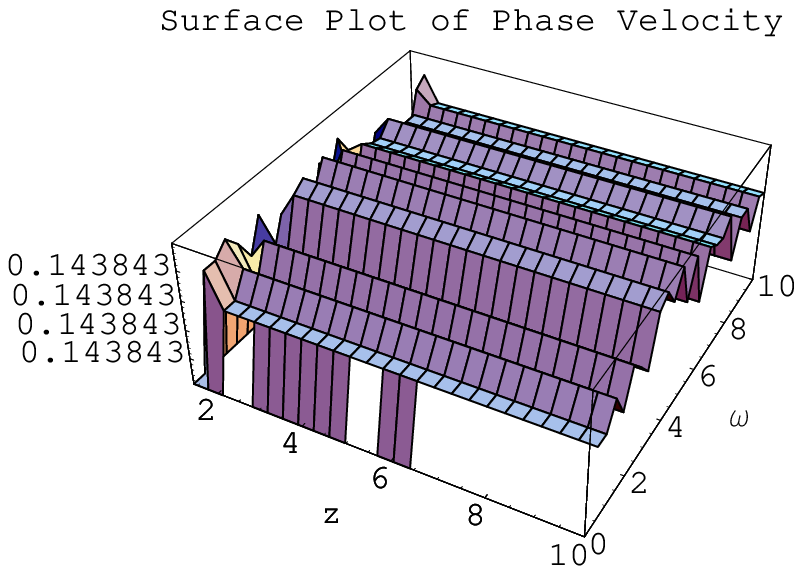,width=0.45\linewidth}
\epsfig{file=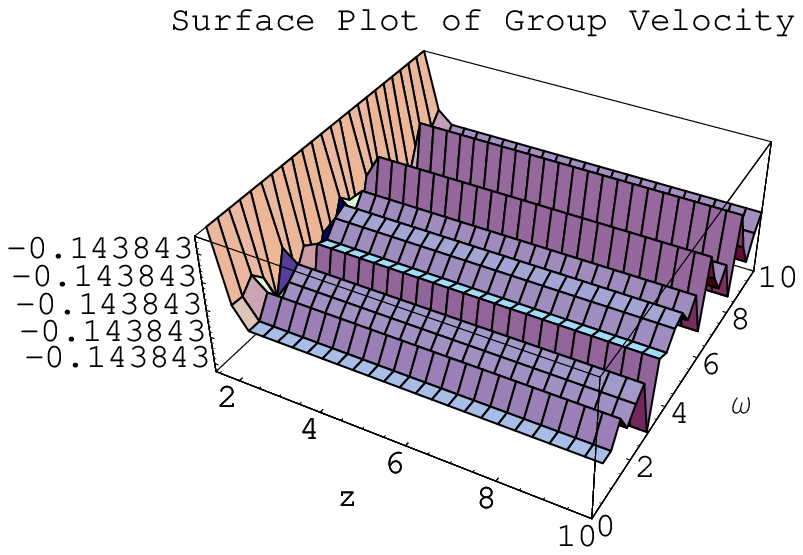,width=0.45\linewidth}\\
\epsfig{file=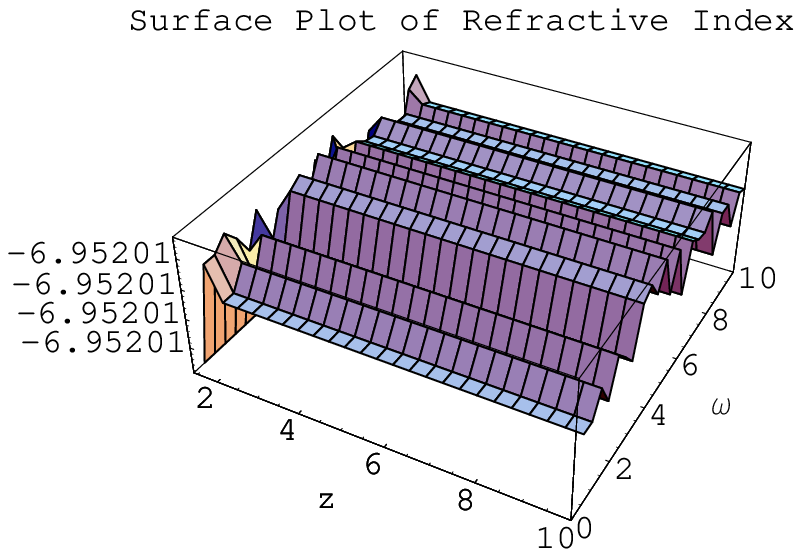,width=0.40\linewidth}
\epsfig{file=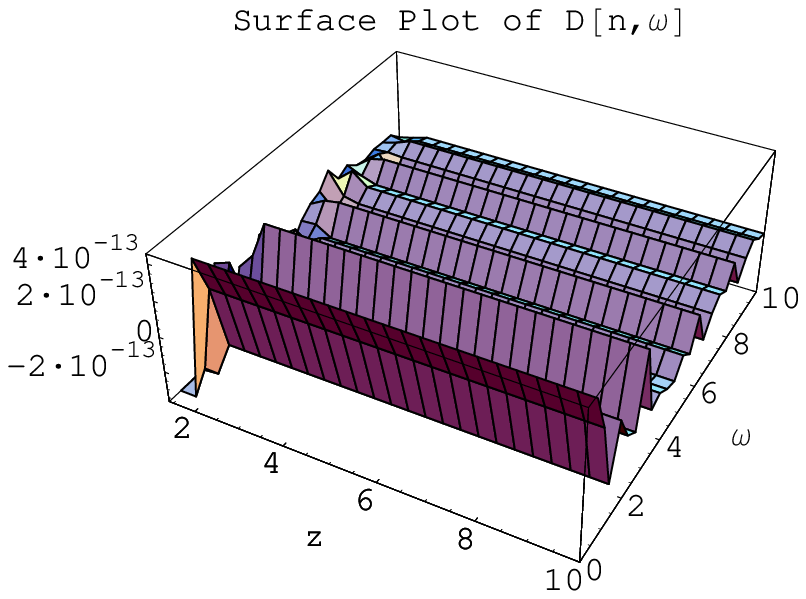,width=0.40\linewidth}
\end{tabular}
\caption{Waves are directed away from the event horizon. Region
has normal and anomalous dispersion at random points.}
\end{figure}
\begin{figure}
\center \epsfig{file=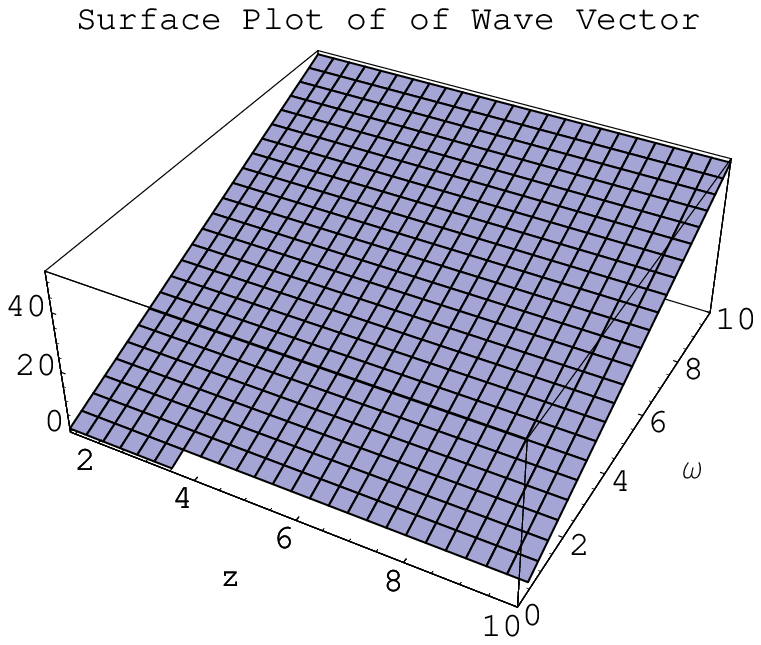,width=0.40\linewidth} \center
\begin{tabular}{cc}
\epsfig{file=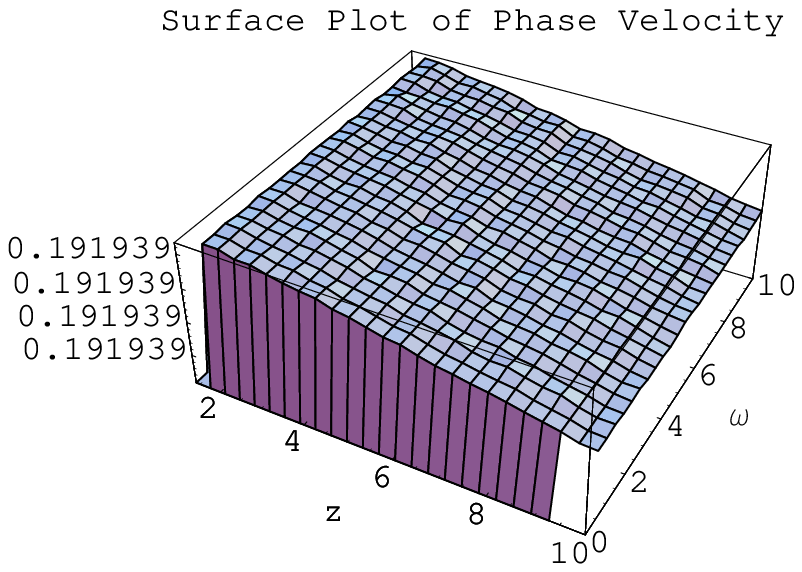,width=0.40\linewidth}
\epsfig{file=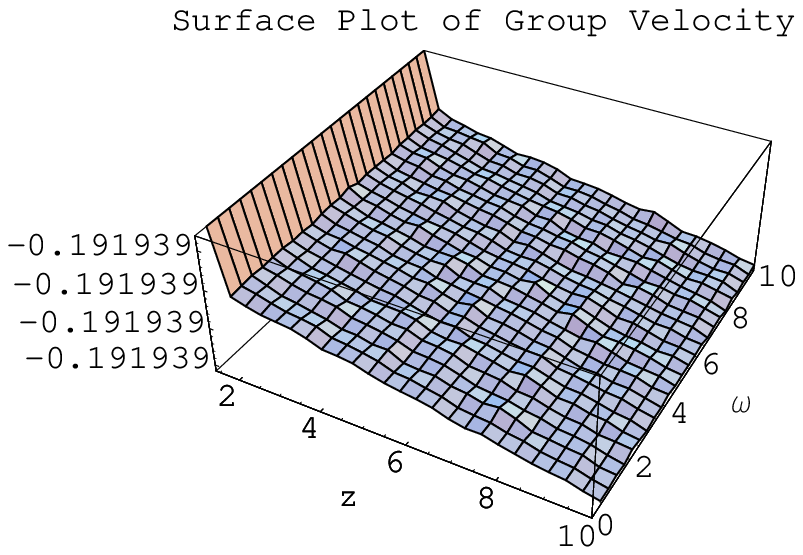,width=0.45\linewidth}\\
\epsfig{file=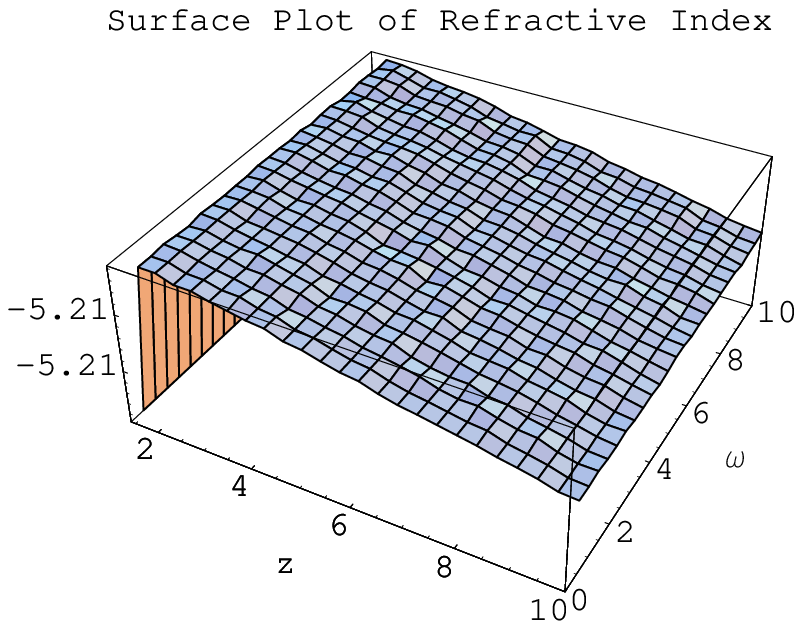,width=0.40\linewidth}
\epsfig{file=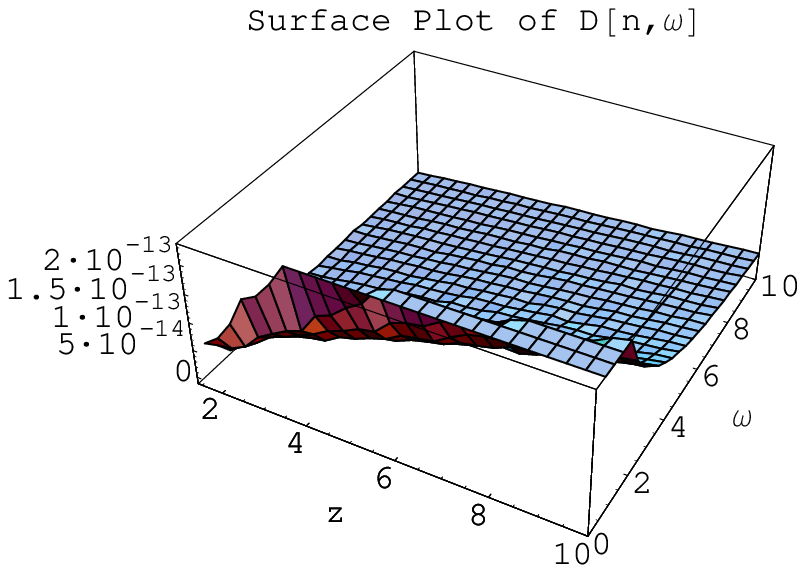,width=0.35\linewidth}
\end{tabular}
\caption{Waves move away from the event horizon. The whole region
has normal dispersion of waves.}
\end{figure}
The results deduced from these figures can be displayed in the
following table.
\begin{center}
Table I. Direction and refractive index of waves
\end{center}
\begin{tabular}{|c|c|c|c|c|}
\hline & \textbf{Direction of Waves} & \textbf{Refractive Index}
($n$)\\ \hline
& & $n<1$ and decreases in the region \\
\textbf{1} & Move towards the event horizon & $1.4\leq z\leq
1.8,0\leq\omega\leq 10$\\&& with the decrease in $z$  \\
\hline
& & $n<1$ and decreases in the region\\
\textbf{2} & Move away from the event horizon & $1.4\leq z\leq
1.6, 0\leq\omega\leq
10$\\&&with the decrease in $z$  \\
\hline
& & $n<1$ and decreases in the region\\
\textbf{3} & Move away from the event horizon & $1.4\leq z\leq1.7,0\leq\omega\leq 10$\\
& &with the decrease in $z$ \\
\hline
\end{tabular}
\\\\
Figures \textbf{1} and \textbf{2} indicate normal and anomalous
dispersion of waves at random points while Figure \textbf{3} gives
normal dispersion in the whole region $1.4\leq
z\leq10,~0\leq\omega\leq 10$. The group and phase velocities are
equal in magnitudes but opposite in directions in all figures.

\section{Rotating Magnetized Flow with Hot Plasma}

In this general case, plasma is assumed to be rotating and
magnetized. Fluid's velocity and magnetic field are supposed to
lie in $xz$-plane. The respective Fourier analyzed perturbed GRMHD
equations, i.e., Eqs.(\ref{25})-(\ref{31}) are given in Section
$3$.

\subsection{Numerical Solutions}

We consider the same assumptions for the values of lapse function,
velocity, pressure, density and specific enthalpy as given in
Section $4$. The restrictions on the magnetic field are as follows
\begin{itemize}
\item $B=\sqrt{\frac{176}{7}}$.
\item For $u=V$ and $V^F=0,$ Eq.(\ref{a}) yields $\lambda=1$.
\end{itemize}
The above restrictions satisfy the perfect GRMHD equations
(Eqs.(\ref{4})-(\ref{8})) for the range $1.4\leq z\leq10,~
0\leq\omega\leq 10$. We have $c_5=0$ from Eqs.(\ref{26})-(\ref{27}).
Using above assumptions in Eqs.(\ref{25}) and (\ref{28})-(\ref{31}),
we obtain two dispersion relations. The real part gives
\begin{equation}{\setcounter{equation}{1}}\label{38}
A_1(z)k^4+A_2(z,\omega)k^3+A_3(z,\omega)k^2+A_4(z,\omega)k+A_5(z,\omega)=0
\end{equation}
yielding four values of $k$ but all are imaginary. The
dispersion relation obtained from imaginary part is
\begin{eqnarray}\label{39}
&&B_1(z)k^5+B_2(z,\omega)k^4+B_3(z,\omega)k^3+B_4(z,\omega)k^2+B_5(z,\omega)k\nonumber\\
&&+B_6(z,\omega)=0
\end{eqnarray}
which gives two real values of $k$. Figures \textbf{4}-\textbf{8}
represent their solutions.\\
The following table shows the results obtained from these figures.
\newpage
\begin{center}
Table II. Direction and refractive index of waves
\end{center}
\begin{tabular}{|c|c|c|c|c|}
\hline & \textbf{Direction of Waves} & \textbf{Refractive Index}
($n$)\\ \hline
& & $n<1$ and increases in the region \\
\textbf{4} & Move towards the event horizon & $1.5\leq z\leq
1.7,0\leq\omega\leq 10$\\&& with the decrease in $z$  \\
\hline
& & $n<1$ and decreases in the region\\
\textbf{5} & Move away from the event horizon & $1.4\leq z\leq
1.8, 0\leq\omega\leq
10$\\&&with the decrease in $z$  \\
\hline
& & $n<1$ and increases in the region\\
\textbf{6} & Move towards the event horizon & $1.4\leq z\leq 1.7,
0\leq\omega\leq
10$\\&&with the decrease in $z$  \\
\hline
& & $n<1$ and increases in the region\\
\textbf{7} & Move away from the event horizon & $1.4\leq z\leq
1.75, 0\leq\omega\leq
10$\\&&with the decrease in $z$  \\
\hline
& & $n<1$ and increases in the region\\
\textbf{8} & Move away from the event horizon & $1.4\leq z\leq
1.8, 0\leq\omega\leq
10$\\&&with the decrease in $z$  \\
\hline
\end{tabular}
\\\\
These figures show that group and phase velocities are
antiparallel.
\begin{figure}
\center \epsfig{file=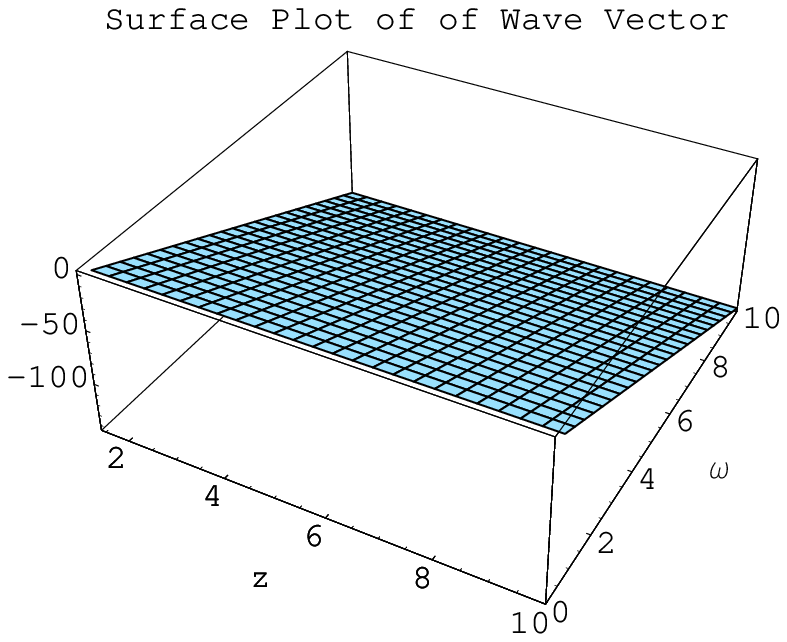,width=0.40\linewidth} \center
\begin{tabular}{cc}
\epsfig{file=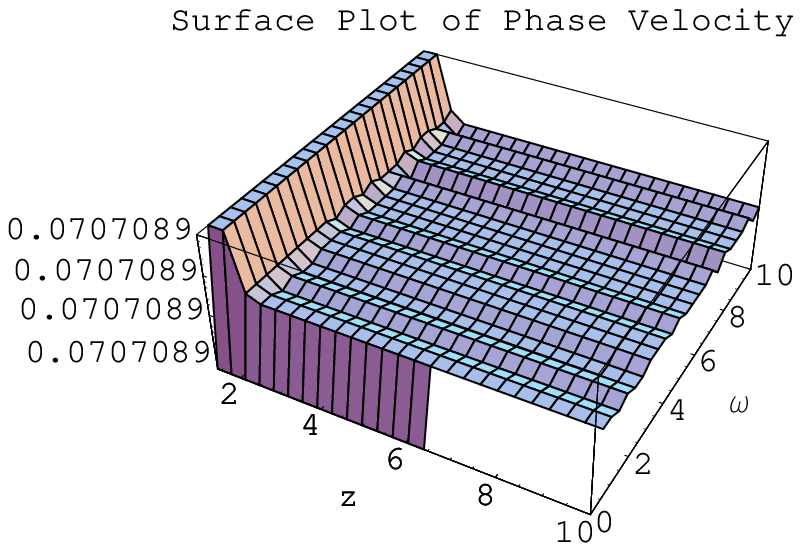,width=0.45\linewidth}
\epsfig{file=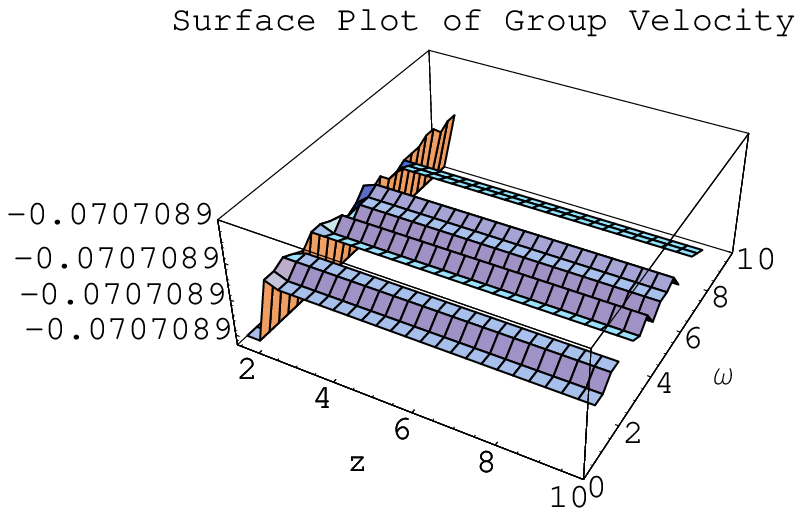,width=0.45\linewidth}\\
\epsfig{file=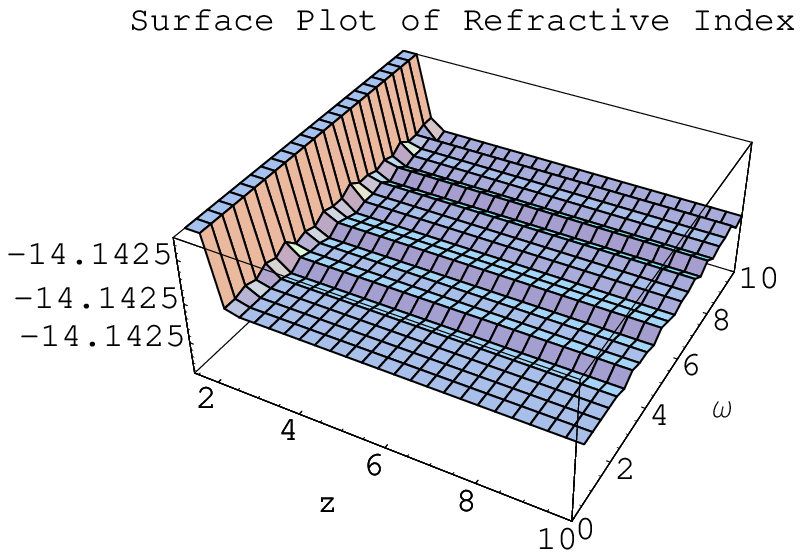,width=0.40\linewidth}
\epsfig{file=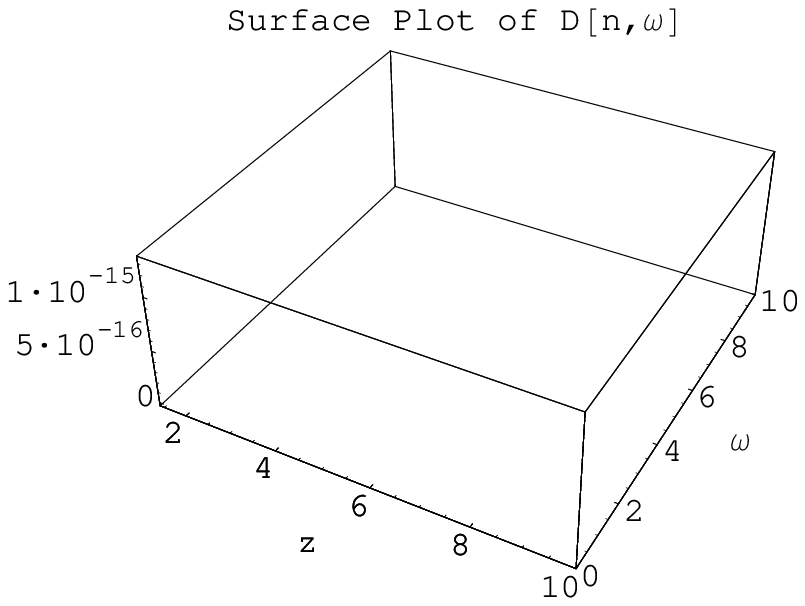,width=0.35\linewidth}
\end{tabular}
\caption{Waves move towards the event horizon. The dispersion is
found to be normal.}
\end{figure}
\begin{figure} \center
\epsfig{file=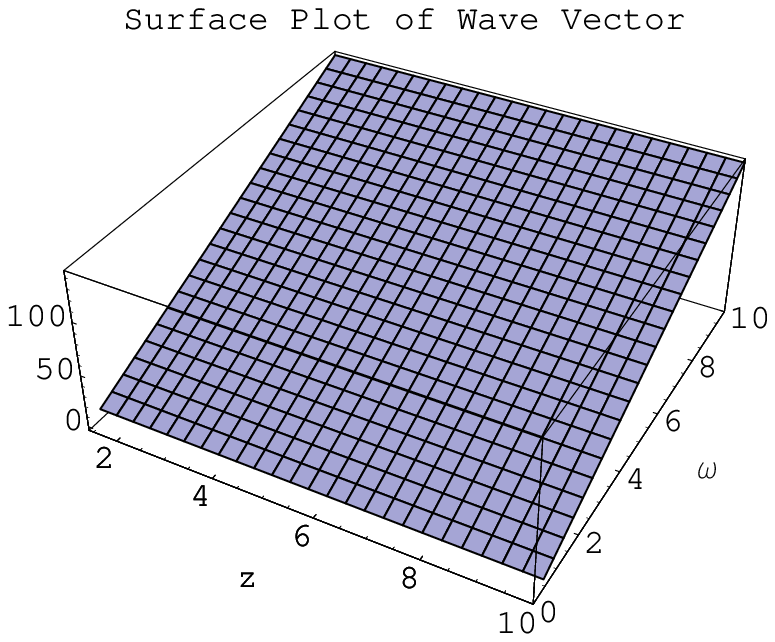,width=0.40\linewidth} \center
\begin{tabular}{cc}
\epsfig{file=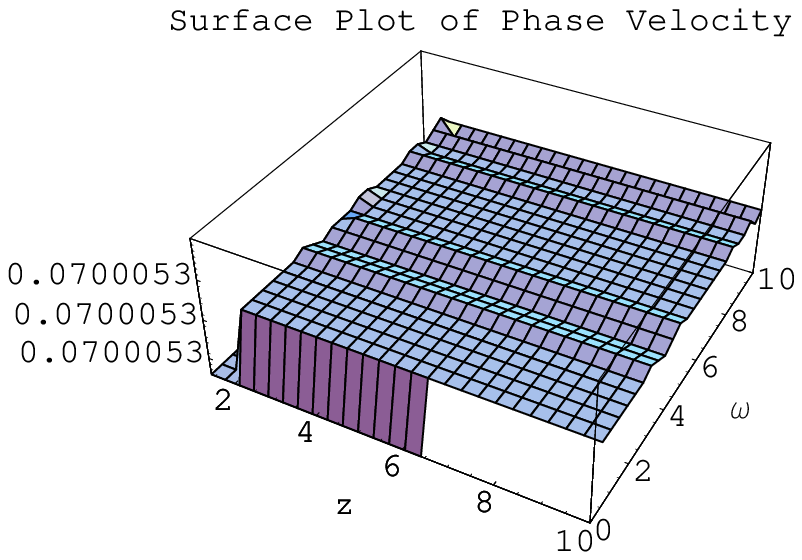,width=0.45\linewidth}
\epsfig{file=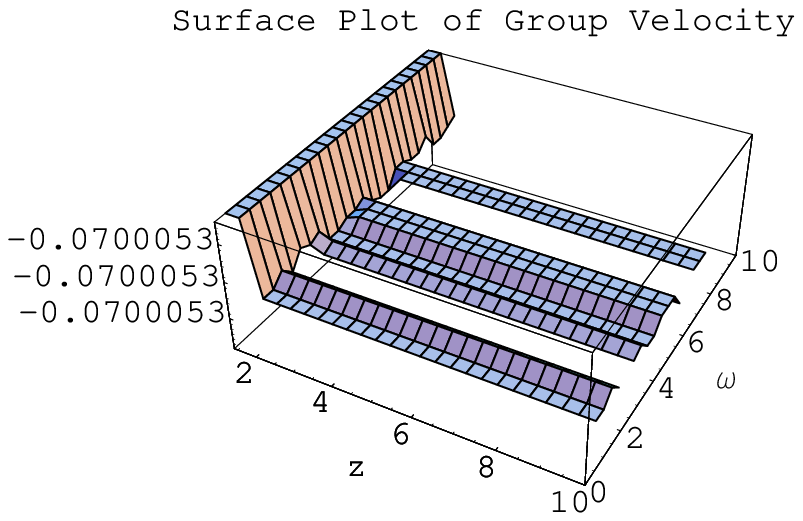,width=0.45\linewidth}\\
\epsfig{file=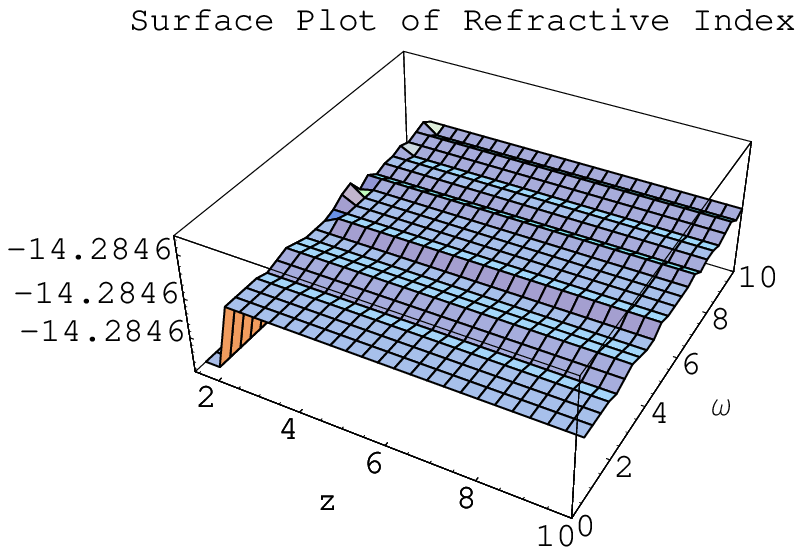,width=0.40\linewidth}
\epsfig{file=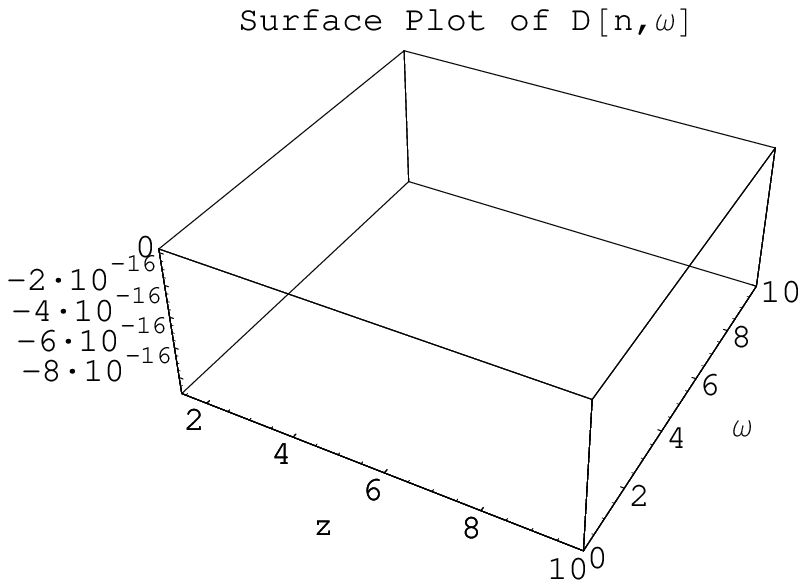,width=0.40\linewidth}
\end{tabular}
\caption{Waves move away from the event horizon. Region has
anomalous dispersion of waves.}
\end{figure}
\begin{figure}
\center \epsfig{file=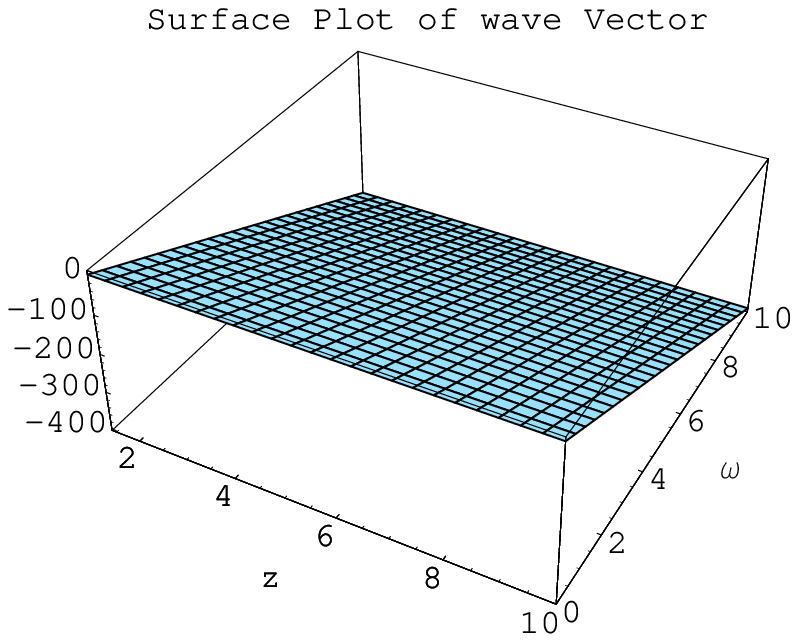,width=0.40\linewidth} \center
\begin{tabular}{cc}
\epsfig{file=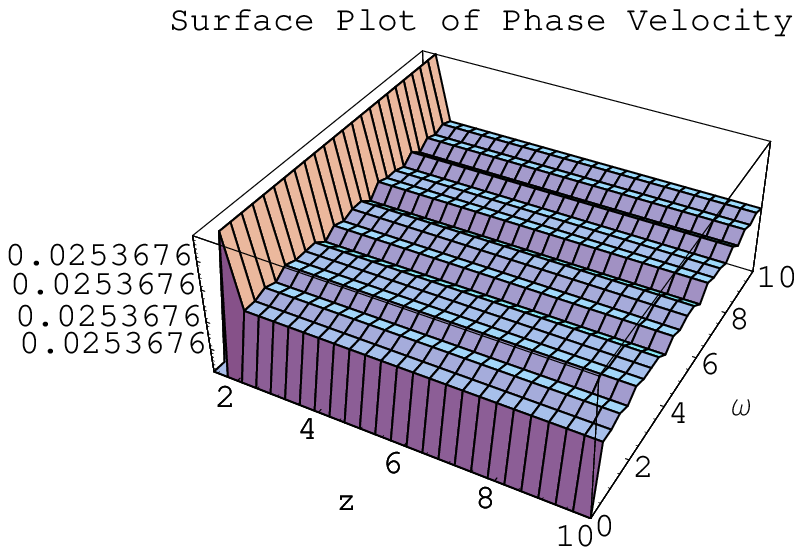,width=0.45\linewidth}
\epsfig{file=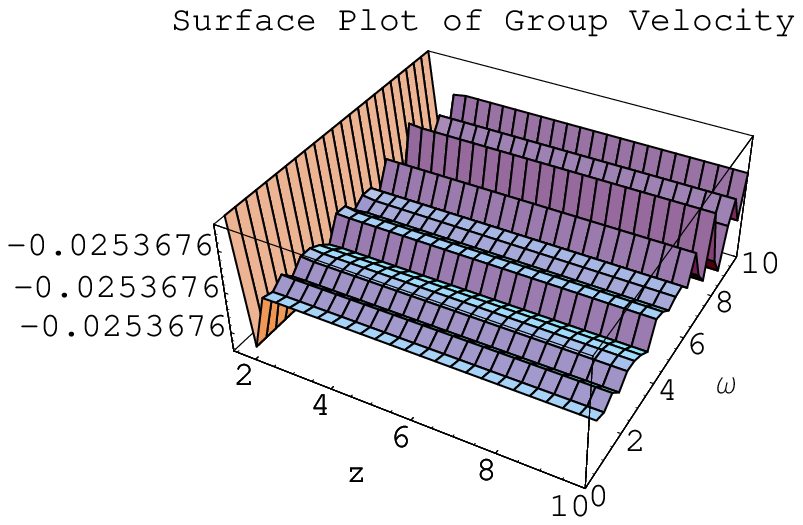,width=0.45\linewidth}\\
\epsfig{file=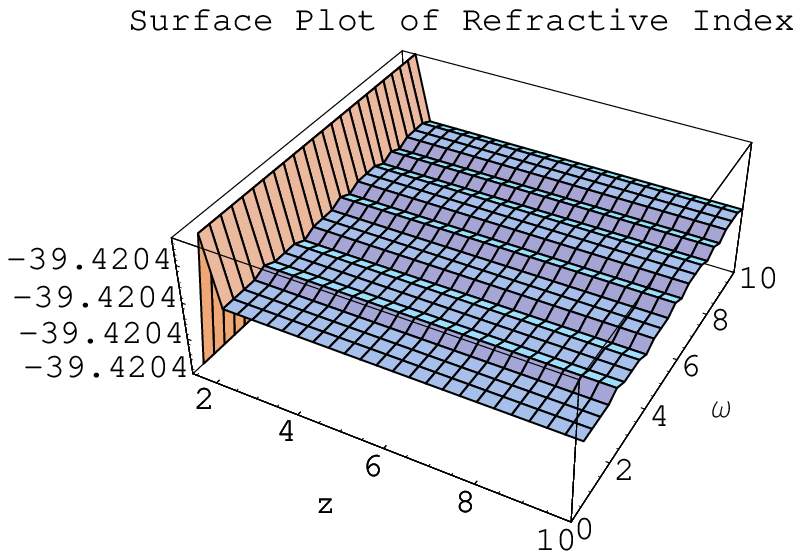,width=0.40\linewidth}
\epsfig{file=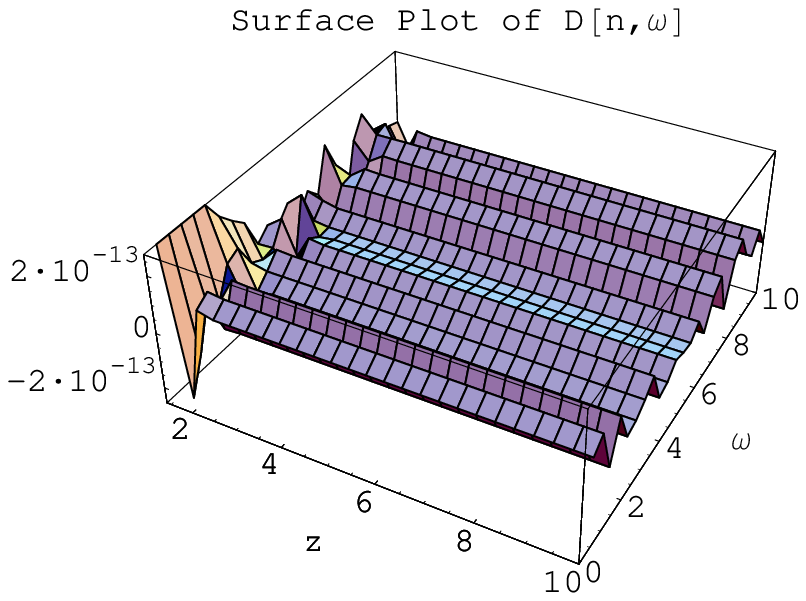,width=0.35\linewidth}
\end{tabular}
\caption{Waves move towards the event horizon. The dispersion is
normal as well as anomalous at random points.}
\end{figure}
\begin{figure}
\center \epsfig{file=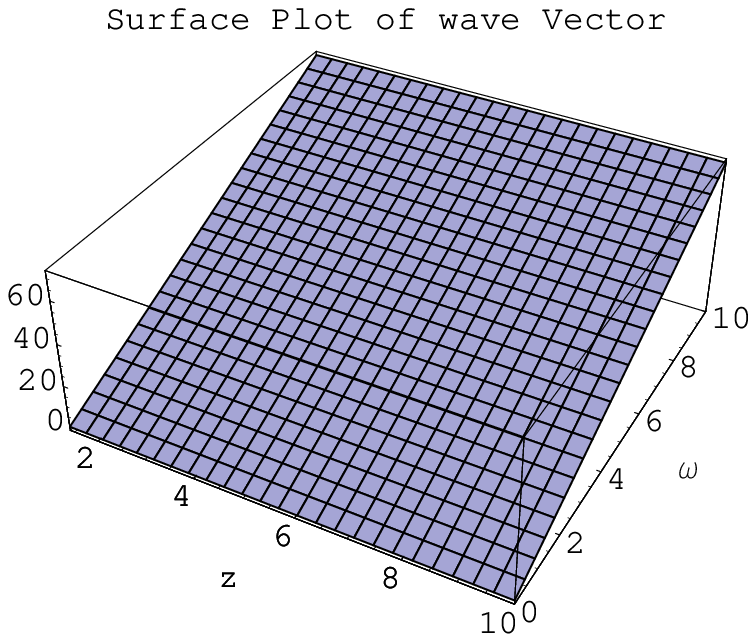,width=0.40\linewidth} \center
\begin{tabular}{cc}
\epsfig{file=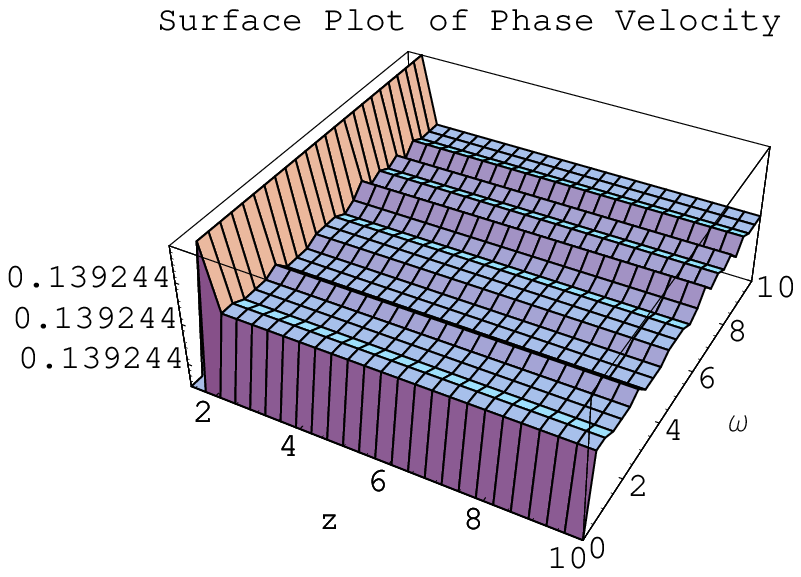,width=0.45\linewidth}
\epsfig{file=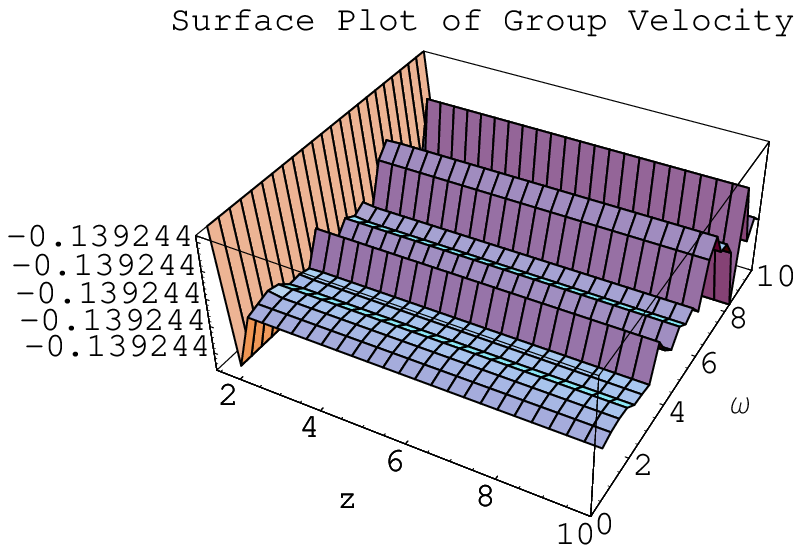,width=0.45\linewidth}\\
\epsfig{file=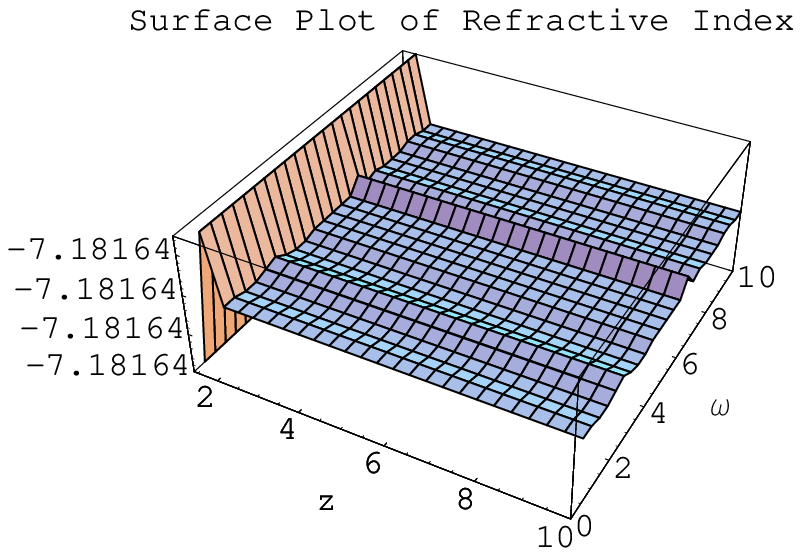,width=0.40\linewidth}
\epsfig{file=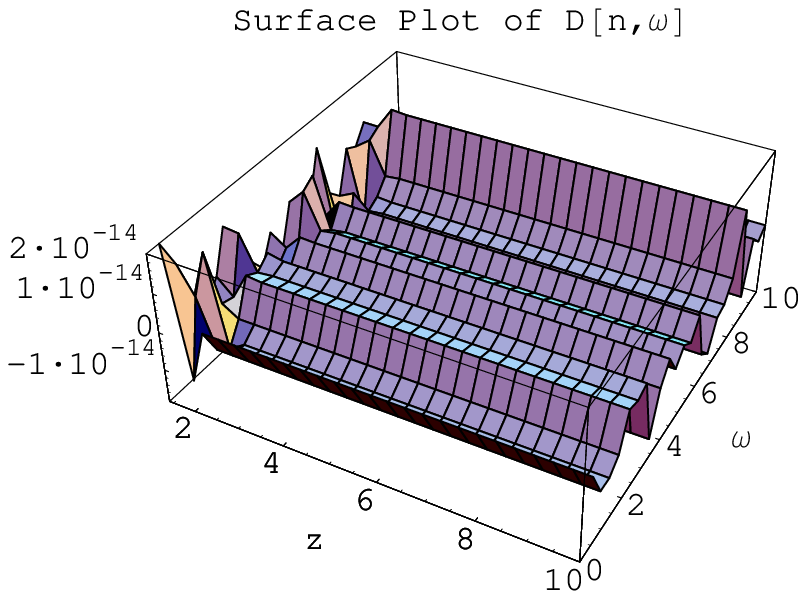,width=0.35\linewidth}
\end{tabular}
\caption{Waves move away from the event horizon. Region has normal
and anomalous dispersion of waves randomly.}
\end{figure}
\begin{figure}
\center \epsfig{file=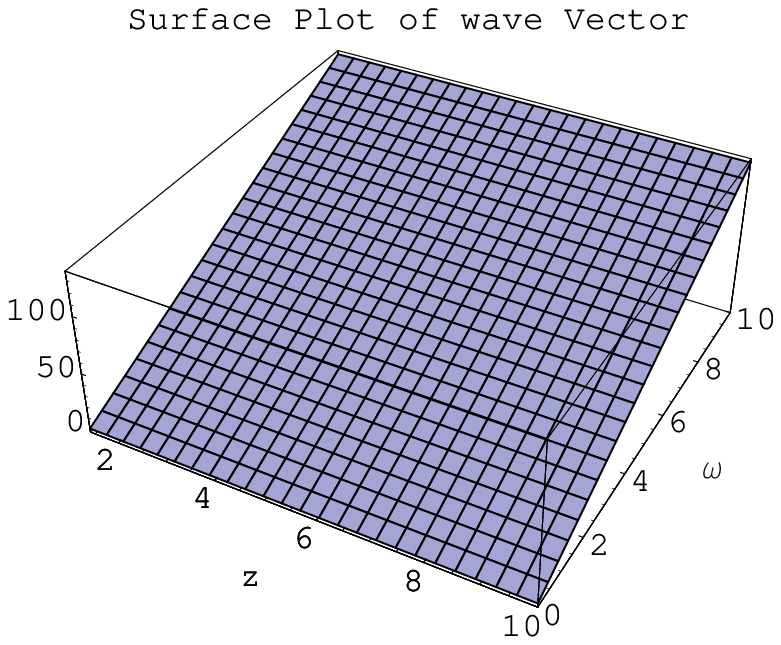,width=0.40\linewidth} \center
\begin{tabular}{cc}
\epsfig{file=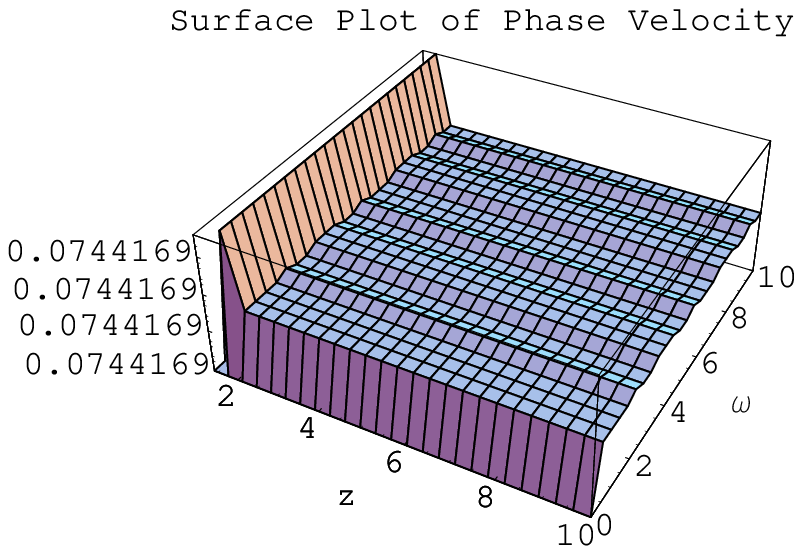,width=0.45\linewidth}
\epsfig{file=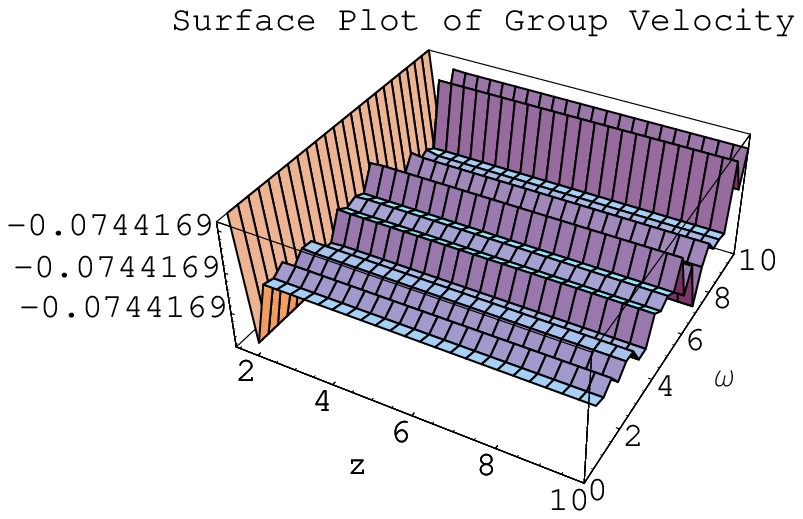,width=0.45\linewidth}\\
\epsfig{file=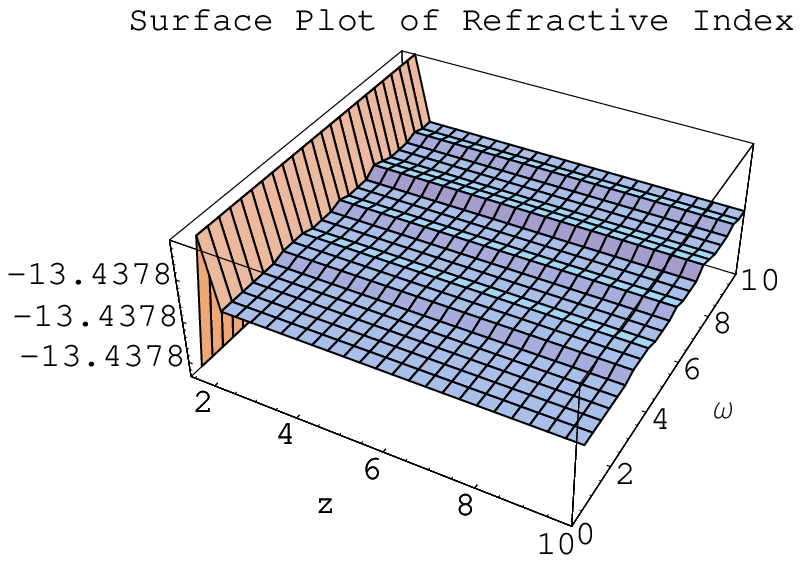,width=0.40\linewidth}
\epsfig{file=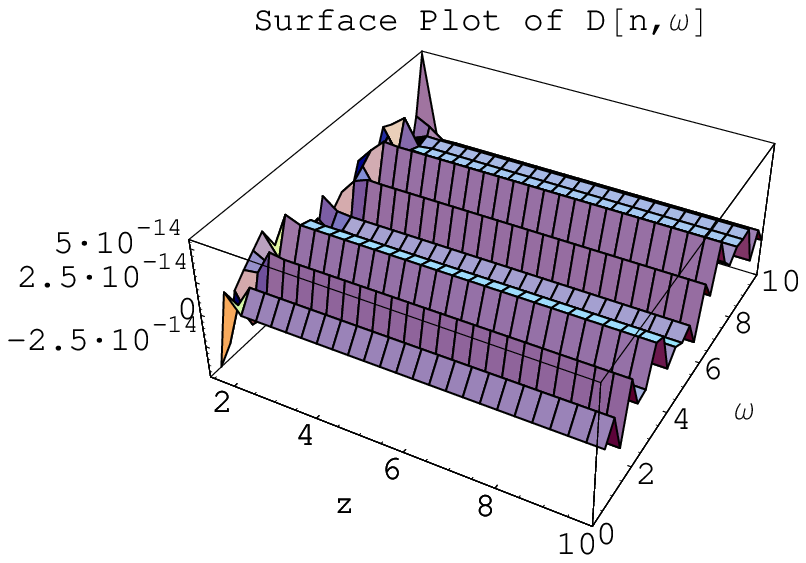,width=0.35\linewidth}
\end{tabular}
\caption{Waves move away from the event horizon. The dispersion is
normal as well as anomalous at random points.}
\end{figure}
\newpage
Normal and anomalous dispersion of waves can be classified into
different regions given in the following table.
\begin{center}
Table III. Regions of dispersion
\end{center}
\begin{center}
\begin{tabular}{|c|c|c|c|c|}
\hline &  \textbf{ Normal dispersion} &  \textbf{Anomalous
dispersion}
\\\hline & $2\leq z\leq 10, 2.9\leq\omega\leq 3$      &
$2\leq z\leq 10, 2.5\leq\omega\leq 2.8$ \\& $2\leq z\leq
10,3.2\leq\omega\leq 3.6$   & $2\leq z\leq 10, 3.7\leq\omega\leq
3.8$
\\& $2\leq z\leq 10, 4.3\leq\omega\leq 4.4$    & $2\leq z\leq
10, 4.1\leq\omega\leq 4.2$  \\& $2\leq z\leq 10, 5.1\leq\omega\leq
5.3$ &
$2\leq z\leq 10, 5.4\leq\omega\leq 5.6$  \\
\textbf{6} & $2\leq z\leq 10, 6.5\leq\omega\leq 6.7$   &
$2.1\leq z\leq 10, 6.2\leq\omega\leq 6.4$    \\
& $2\leq z\leq 10, 7.1\leq\omega\leq 7.3$     &
$2\leq z\leq 10, 7.4\leq\omega\leq 7.6$\\
& $2\leq z\leq 10, 8.4\leq\omega\leq 8.6$      &
$2\leq z\leq 10, 8.1\leq\omega\leq 8.2$   \\
& $2\leq z\leq 10, 9.4\leq\omega\leq 9.6$     & $2\leq z\leq 10,
9\leq\omega\leq 9.2$
\\\hline
& $2\leq z\leq 10, 2\leq\omega\leq 2.2$  &
$2\leq z\leq 10, 2.6\leq\omega\leq 2.7$     \\
& $2\leq z\leq 10, 3.5\leq\omega\leq 3.6$  &
$2\leq z\leq 10, 3.2\leq\omega\leq 3.4$     \\
& $2\leq z\leq 10, 4.1\leq\omega\leq 4.3$   &
$1.6\leq z\leq 10, 4.4\leq\omega\leq 4.5$      \\
& $1.6\leq z\leq 10, 5.1\leq\omega\leq 5.2$  &
$1.6\leq z\leq 10, 5.3\leq\omega\leq 5.4$  \\
& $2\leq z\leq 10, 6\leq\omega\leq 6.2$   & $1.6\leq z\leq
10,6.6\leq\omega\leq 6.8$\\
\textbf{7}& $2\leq z\leq 10, 7.7\leq\omega\leq 7.8$   &
$2\leq z\leq 10, 7.1\leq\omega\leq 7.2$  \\
& $2\leq z\leq 10, 8.3\leq\omega\leq 8.6$   & $2\leq z\leq 10,
8\leq\omega\leq 8.2$  \\
& $2\leq z\leq 10, 9.2\leq\omega\leq 9.4$   & $2\leq z\leq 10,
9.5\leq\omega\leq 9.9$
\\\hline & $1.5\leq z\leq 10, 2\leq\omega\leq 2.2$      &
$1.5\leq z\leq 10, 2.6\leq\omega\leq 2.8$ \\& $1.5\leq z\leq
10,3.5\leq\omega\leq 3.6$   & $1.5\leq z\leq 10, 3.2\leq\omega\leq
3.3$
\\& $1.6\leq z\leq 10, 4\leq\omega\leq 4.2$    & $1.6\leq z\leq
10, 4.4\leq\omega\leq 4.5$  \\& $1.8\leq z\leq 10,
5.6\leq\omega\leq 5.7$ &
$1.8\leq z\leq 10, 5.2\leq\omega\leq 5.4$  \\
\textbf{8} & $2\leq z\leq 10, 6\leq\omega\leq 6.2$   &
$2.1\leq z\leq 10, 6.3\leq\omega\leq 6.4$    \\
& $2\leq z\leq 10, 7.7\leq\omega\leq 7.8$     &
$2\leq z\leq 10, 7.1\leq\omega\leq 7.2$\\
& $2\leq z\leq 10, 8.3\leq\omega\leq 8.6$      &
$2\leq z\leq 10, 8\leq\omega\leq 8.2$   \\
& $2\leq z\leq 10, 9.2\leq\omega\leq 9.4$     & $2\leq z\leq 10,
9.5\leq\omega\leq 9.9$
\\\hline
\end{tabular}
\end{center}

\section{Summary}

In this paper, wave properties of hot plasma in the vicinity of the
Schwarzschild event horizon for a Veselago medium are analyzed. The
$3+1$ GRMHD equations are reformulated for this unusual medium.
Linear perturbation in $3+1$ perfect GRMHD equations is considered
and their component form is derived. Dispersion relations are found
by using Fourier analysis technique for the rotating non-magnetized
and rotating magnetized plasmas.

In the rotating non-magnetized background, Figure \textbf{1} shows
that waves move towards the event horizon while Figures \textbf{2}
and \textbf{3} indicate that waves are directed away from the event
horizon. Dispersion is found to be normal and anomalous at random
points in Figures \textbf{1} and \textbf{2} while it is normal in
most of the region in Figure \textbf{3}.

For the rotating magnetized plasma, Figures \textbf{4} and
\textbf{6} indicate that waves are directed towards the event
horizon while waves move away from the event horizon in Figures
\textbf{5}, \textbf{7} and \textbf{8}. Dispersion is normal in
Figure \textbf{4} while it is anomalous in Figure \textbf{5}.
Dispersion is normal as well as anomalous randomly in Figures
\textbf{6}, \textbf{7} and \textbf{8}. The value of refractive index
is less than $1$ and also phase and group velocities are
antiparallel in all the figures which are the significant features
of this unusual medium. Thus, the presence of Veselago medium is
confirmed for both rotating (non-magnetized and magnetized) plasmas.

The comparison of the results for isothermal \cite{33} and hot
plasma can be summarized in the following table.
\begin{center}
Table IV. Comparison of the results
\end{center}
\begin{tabular}{|c|c|c|}
\hline
\textbf{Results} & \textbf{Isothermal Plasma} & \textbf{Hot Plasma} \\
\hline & & \\
\textbf{Existence of waves} & No waves in rotating & Waves exist in rotating\\
& magnetized plasma & magnetized plasma \\\hline
\textbf{Direction of waves} & Some waves move away  & Most of the waves move\\
& from the event horizon & away from the event horizon\\\hline
\textbf{Dispersion} & Normal at random points & Normal in most of the  \\
&& region in Figures 3 and 4
\\\hline
\end{tabular}\\\\

The difference between our work and previous work is that we have
used the variable specific enthalpy and previous work has been done
using constant specific enthalpy. In our work, waves exist in
rotating magnetized plasma while in previous work, there does not
exist any wave in rotating magnetized plasma. Also, we have found
that most of the waves move away from the event horizon while for
isothermal case, some waves move away from the event horizon. This
comparison shows that variation in specific enthalpy effects the
direction of waves. Dispersion is normal at random points for
isothermal while for hot plasma it is normal in most of the region
in Figures \textbf{3} and \textbf{4}. It is interesting to mention
here that the properties of a Veselago medium turn out for both
rotating (non-magnetized and magnetized) plasmas confirming its
validity.
\renewcommand{\theequation}{A\arabic{equation}}

\section*{Appendix A}

This Appendix contains the Maxwell equations, the GRMHD equations
for the general line element and the Schwarzschild planar analogue
in a Veselago medium ($\epsilon<0,~\mu<0$). In this medium, the
Maxwell equations are
\begin{eqnarray}{\setcounter{equation}{1}}
\label{40}&&\nabla.\textbf{B}=0,\\
\label{41}&&\nabla\times\textbf{E}+\frac{\partial\textbf{B}}{\partial
t}=0,\\
\label{42}&&\nabla\cdot\textbf{E}=-\frac{\rho_e}{\epsilon},\\
\label{43}&&\nabla\times\textbf{B}=-\mu\textbf{j}+\frac{\partial\textbf{E}}{\partial
t}=0.
\end{eqnarray}
Also, the GRMHD equations take the form
\begin{eqnarray}\label{44}
&&\frac{d\textbf{B}}{d\tau}
+\frac{1}{\alpha}(\textbf{B}.\nabla)\beta
+\theta\textbf{B}=-\frac{1}{\alpha}\nabla\times(\alpha\textbf{V}\times\textbf{B}),\\
\label{45}
&&\nabla.\textbf{B}=0,\\
\label{46}
&&\frac{D\rho_0}{D\tau}+\rho_0\gamma^2\textbf{V}.\frac{D\textbf{V}}{D\tau}
+\frac{\rho_0}{\alpha}\left\{\frac{g,_t}{2g}+\nabla.(\alpha\textbf{V}-\beta)\right\}=0,\\\label{47}
&&\left\{\left(\rho_0\mu\gamma^2+\frac{\textbf{B}^2}{4\pi}\right)\gamma_{ij}
+\rho_0\mu\gamma^4V_iV_j
-\frac{1}{4\pi}B_iB_j\right\}\frac{DV^j}{D\tau}
\nonumber\\
&&+\rho_0\gamma^2V_i\frac{D\mu}{D\tau}
-\left(\frac{\textbf{B}^2}{4\pi}\gamma_{ij}-\frac{1}{4\pi}B_iB_j\right)
V^j_{|k}V^k=-\rho_0\gamma^2\mu\{a_i\nonumber\\
&&-\frac{1}{\alpha}\beta_{j|i}V^j -(\pounds_t\gamma_{ij})V^j\}
-p_{|i}+\frac{1}{4\pi}(\textbf{V}\times
\textbf{B})_i\nabla.(\textbf{V}\times\textbf{B})\nonumber\\
&&-\frac{1}{8\pi\alpha^2}(\alpha\textbf{B})^2_{|i}+\frac{1}{4\pi\alpha}(\alpha
B_i)_{|j}B^j-\frac{1}{4\pi\alpha}(\textbf{B}\times\{\textbf{V}\times
[\nabla\nonumber\\
&&\times(\alpha\textbf{V}\times\textbf{B})
-(\textbf{B}.\nabla)\beta]+(\textbf{V}\times\textbf{B}).\nabla\beta\})_i,\\
\label{48}&&\frac{D}{D\tau}(\mu\rho_0\gamma^2)-\frac{d
p}{d\tau}+\Theta(\mu\rho_0\gamma^2-p)+\frac{1}{2\alpha}
(\mu\rho_0\gamma^2V^{i}V^{j}\nonumber\\&&+p\gamma_{ij})
\pounds_t\gamma_{ij}+2\mu\rho_0\gamma^2(\textbf{V}.\textbf{a})+\mu\rho_0\gamma^2(\nabla.\textbf{V})
-\frac{1}{\alpha}\beta^{j,i}\nonumber\\
&&\times(\mu\rho_0\gamma^2V_{i}V_{j}+p\gamma_{ij})-\frac{1}{4\pi}
(\textbf{V}\times\textbf{B}).(\textbf{V}\times\frac{d\textbf{B}}{d\tau})
-\frac{1}{4\pi}\nonumber\\ &&\times
(\textbf{V}\times\textbf{B}).(\textbf{B}\times\frac{d\textbf{V}}{d\tau})-\frac{1}{4\pi\alpha}
(\textbf{V}\times\textbf{B}).\nabla\beta-\frac{1}{4\pi}\theta(\textbf{V}\times\textbf{B})
\nonumber\\&&.(\textbf{V}\times\textbf{B})+\frac{1}{4\pi\alpha}(\textbf{V}\times\textbf{B})
.(\nabla\times\alpha\textbf{B})=0.
\end{eqnarray}
For the Schwarzschild planar analogue, $\beta,~\theta$ and
$\pounds_t\gamma_{ij}$ vanish, the perfect GRMHD equations become
\begin{eqnarray}\label{49}
&&\frac{\partial\textbf{B}}{\partial t}=-\nabla \times(\alpha
\textbf{V}\times\textbf{B}),\\\label{50}
&&\nabla.\textbf{B}=0,\\\label{51}
&&\frac{\partial\rho_0}{\partial
t}+(\alpha\textbf{V}.\nabla)\rho_0+\rho_0\gamma^2
\textbf{V}.\frac{\partial\textbf{V}}{\partial
t}+\rho_0\gamma^2\textbf{V}.(\alpha\textbf{V}.\nabla)\textbf{V}\nonumber\\
&&+\rho_0{\nabla.(\alpha\textbf{V})}=0, \\\label{52}
&&\{(\rho_0\mu\gamma^2+\frac{\textbf{B}^2}{4\pi})\delta_{ij}
+\rho_0\mu\gamma^4V_iV_j
-\frac{1}{4\pi}B_iB_j\}(\frac{1}{\alpha}\frac{\partial}{\partial
t}+\textbf{V}.\nabla)V^j\nonumber
\end{eqnarray}
\begin{eqnarray}
&&-(\frac{\textbf{B}^2}{4\pi}\delta_{ij}-\frac{1}{4\pi}B_iB_j)
V^j,_kV^k+\rho_0\gamma^2V_i\{\frac{1}{\alpha}\frac{\partial
\mu}{\partial t}+(\textbf{V}.\nabla)\mu\}\nonumber\\
&&=-\rho_0\mu\gamma^2a_i-p,_i+
\frac{1}{4\pi}(\textbf{V}\times\textbf{B})_i\nabla.(\textbf{V}\times\textbf{B})
-\frac{1}{8\pi\alpha^2}(\alpha\textbf{B})^2,_i\nonumber\\
&&+\frac{1}{4\pi\alpha}(\alpha B_i),_jB^j-\frac{1}{4\pi\alpha}
[\textbf{B}\times\{\textbf{V}\times(\nabla\times(\alpha\textbf{V}
\times\textbf{B}))\}]_i,\\
\label{53}&&(\frac{1}{\alpha}\frac{\partial}{\partial
t}+\textbf{V}.\nabla)(\mu\rho_0\gamma^2)-\frac{1}{\alpha}\frac{\partial
p }{\partial
t}+2\mu\rho_0\gamma^2(\textbf{V}.\textbf{a})+\mu\rho_0\gamma^2
(\nabla.\textbf{V})\nonumber\\&&-\frac{1}{4\pi}
(\textbf{V}\times\textbf{B}).(\textbf{V}\times\frac{1}{\alpha}\frac{\partial
\textbf{B}}{\partial t})-\frac{1}{4\pi}
(\textbf{V}\times\textbf{B}).(\textbf{B}\times\frac{1}{\alpha}\frac{\partial
\textbf{V}}{\partial
t})\nonumber\\&&+\frac{1}{4\pi\alpha}(\textbf{V}\times\textbf{B}).
(\nabla\times\alpha\textbf{B})=0.
\end{eqnarray}

\end{document}